\documentclass[a4paper,11pt]{article}
\pdfoutput=1 
\usepackage{jheppub}
\usepackage[T1]{fontenc}
\usepackage{multirow}
\usepackage{booktabs}
\usepackage[usenames,dvipsnames,svgnames,table]{xcolor}

\definecolor{interesting}{rgb}{0., 1., 0.5}
\definecolor{suppressed}{rgb}{0.97, 0.51, 0.47}

\newcommand{\sigmav}{\ensuremath{\langle \sigma v \rangle}}
\let\oldcite\cite
\renewcommand{\cite}{~\oldcite}
\newcommand{\tev}{\ensuremath{\,\text{TeV}}}
\newcommand{\gev}{\ensuremath{\,\text{GeV}}}
\newcommand{\cmCubePerSec}{\ensuremath{\,\text{cm}^3/\text{s}}}
\title{Model-independent analysis of the DAMPE excess}

\author[a]{Peter Athron\note{Corresponding author.}}
\author[a,b]{Csaba Balazs}
\author[a]{Andrew Fowlie}
\author[a,1]{Yang Zhang}

\affiliation[a]{ARC Centre of Excellence for Particle Physics at the Tera-scale, School of Physics and Astronomy, Monash University, Melbourne, Victoria 3800, Australia}
\affiliation[b]{Monash Centre for Astrophysics, School of Physics and Astronomy, Monash University, Melbourne, Victoria 3800 Australia}

\emailAdd{peter.athron@coepp.org.au}
\emailAdd{csaba.balazs@monash.edu}
\emailAdd{andrew.fowlie@monash.edu}
\emailAdd{yang.zhang@monash.edu}

\abstract{The Dark Matter Particle Explorer (DAMPE) recently released measurements of the electron spectrum with a hint of a narrow peak at about $1.4\tev$. We investigate dark matter (DM) models that could produce such a signal by annihilation in a nearby subhalo whilst simultaneously satisfying constraints from DM searches. In our model-independent approach, we consider all renormalizable interactions via a spin 0 or 1 mediator between spin 0 or 1/2 DM particles and the Standard Model leptons. We find that of the 20 combinations, 10 are ruled out by velocity or helicity suppression of the annihilation cross section to fermions. The remaining 10 models, though, evade constraints from the relic density, collider and direct detection searches, and include models of spin 0 and 1/2 DM coupling to a spin 0 or 1 mediator. We delineate the regions of mediator mass and couplings that could explain the DAMPE excess. In all cases the mediator is required to be heaver than about $2\tev$ by LEP limits.}

\begin{document} 
\maketitle
\flushbottom

\section{Introduction}\label{sec:intro}

The Dark Matter Particle Explorer (DAMPE)\cite{TheDAMPE:2017dtc} recently measured the cosmic ray spectrum of high energy electrons and positrons\cite{Ambrosi:2017wek}. These electrons and positrons are an important potential probe of new physics such as dark matter (DM) decay or annihilation within the Milky Way galaxy. The DAMPE measurement of the energy spectrum extends previous direct measurements up to about 5 TeV.  One of the most exciting aspects of the DAMPE measurements is the confirmation of a spectral break somewhat below $1\tev$. The other exciting feature is due to the excellent energy resolution of DAMPE: there appears to be a sharp resonant feature in the data at about $1.4\tev$.  This is despite the admittedly sizeable statistical and systematic uncertainties. 

While this feature could be a statistical fluctuation\cite{Fowlie:2017fya} or may be due to standard astrophysical sources, it could also be the first precursor of dark matter  detection.  The sharp peak observed in the positron spectrum could originate from the annihilation of dark matter into electrons.  Here we investigate the viability of particle dark matter models as an explanation for this excess. We  classify the new states and interactions that can explain the excess, constructing a collection of simplified models, and check in each case whether the signal can be achieved, while simultaneously fitting the relic density and evading a variety of relevant experimental constraints that we identify. 

To describe the excess of electrons, we assume that DM is leptophilic\cite{Bell:2014tta,DEramo:2017zqw,Fox:2008kb,Cohen:2009fz,Schmidt:2012yg,Ibarra:2009bm,Bi:2009uj,Kopp:2014tsa,Kyae:2009jt,Chun:2009zx,Spolyar:2009kx,Agrawal:2014ufa,Haba:2010ag,Farzan:2010mr,Ko:2010at,Boucenna:2015tra,Freitas:2014jla,Cao:2014cda,Das:2013jca,Davoudiasl:2009dg,Kopp:2010su,Cavasonza:2016qem,Carone:2011iw,Chen:2015tia,Buckley:2015cia,Dev:2013hka,Dev:2016qbd}, that is, at tree-level it couples only to leptons and possibly neutrinos.  This is motivated by simplicity and since it ameliorates constraints from searches for DM in proton collisions at the LHC and searches for elastic scatters between DM and nuclei in direct detection (DD) experiments.  Our DM candidate is a Weakly Interacting Massive Particle (WIMP) that satisfies the required abundance of DM with an annihilation cross section at thermal freezeout of about $\sigmav \approx 10^{-26} \cmCubePerSec$.

Self-annihilation of the dark matter particles could result in the injection of a resonant source, $\delta(E - m_\chi)$, of charged particles in the Milky Way.  The sharp injection spectrum from such a source, however, is softened and smeared by diffusion in Galactic magnetic fields, bremsstrahlung and inverse Compton scattering with CMB photons.  The spectrum can be described by a diffusion and loss equation,
\begin{equation}
    \frac{\partial \rho(t, x, E)}{\partial t} = \vec\nabla \cdot \left[d(E)\vec\nabla \rho(t, x, E)\right] + \frac{\partial \left[l(E) \rho(t, x, E)\right]}{\partial E} + J(t, x, E),
\end{equation}
where $\rho(t, x, E)$ is the energy spectrum, $d(E)$ is a diffusion coefficient, $l(E)$ is an energy loss coefficient, and $J(t, x, E)$ is a source.  Assuming that energy losses dominate diffusion and a steady-state spectrum, for a resonance source, $J(t, x, E) = J_0 \delta(E - m_\chi)$, the energy spectrum exhibits an endpoint,
\begin{equation}
    \rho(E) = 
    \begin{cases} 
      \frac{J_0}{l(E)} & E \le m_\chi\\
      0 & \text{otherwise,}\\
   \end{cases}
\end{equation}
rather than a sharp resonance, such as the source\cite{Slatyer:2017sev}. We thus require a nearby and late-time injection of monochromatic electrons, such that losses do not dominate. We thus assume a nearby source of DM annihilation, such as a subhalo within a few kpc\cite{Yuan:2017ysv,Fan:2017sor,Fang:2017tvj}.

After kinetic decoupling, DM forms gravitationally bound subhalo structures.  These structures predominantly merge into a DM halo, but subhalos may survive.  Indeed, numerical $n$-body simulations\cite{Diemand:2008in} predict numerous subhalo structures in a halo comparable to that of the Milky Way.  A signal from a subhalo with an annihilation cross section $\sigmav \approx \sigmav_{v\to0} \approx 10^{-26} \cmCubePerSec$ could be enhanced by a substantial DM density, reaching the observed amplitude of the DAMPE signal. Because electrons from a local subhalo would, nevertheless, lose energy, we assume a DM mass slightly greater than the peak observed by DAMPE in the electron energy spectrum, $m_\chi \approx 1.5\tev$.

The result immediately generated considerable interest\cite{Niu:2017hqe,Gao:2017pym,Jin:2017qcv,Huang:2017egk,Duan:2017qwj,Gu:2017bdw,Chao:2017yjg,Tang:2017lfb,Zu:2017dzm,Liu:2017rgs,Cao:2017ydw,Gu:2017gle,Duan:2017pkq,Fang:2017tvj,Fan:2017sor,Yuan:2017ysv} and particle dark matter interpretations have already been investigated. 
In\cite{Yuan:2017ysv} they consider lepton portal, lepton flavour and TeV right-handed neutrino models. In\cite{Fan:2017sor} they consider a specific model that can explain both neutrino masses and the DAMPE excess, where the DM candidate is a vector-like fermion, with a $Z^\prime$ mediator.  In\cite{Gu:2017gle} they propose an electrophilic dark matter candidate, which may be a fermion with scalar mediator or the dark matter candidate could be a scalar with a fermion mediator but perform calculations only for the former. Finally in\cite{Duan:2017pkq} they consider simplified models that can explain the excess where the DM candidate is a leptophilic fermion.

Our work, presented in this paper, is closest in spirit to\cite{Duan:2017pkq}, however we are considerably more general.  We present a set of simplified leptophilic models with scalar, Majorana and Dirac fermions, that couple with either scalar or vector mediators.  For the scalar mediator we consider both scalar and pseudoscalar couplings to fermions and for the vector mediator we consider vector and axial vector couplings to fermions, while for scalar dark matter we consider cubic and quartic interactions with the mediator. For simplicity we neglect more exotic cases; we do not consider e.g., spin-1 DM, a spin-1/2 $t$-channel mediator or a spin-2 mediator in this work. We systematically investigate each case to determine if scenarios that fit the relic density can explain the DAMPE excess and if so, whether or not those scenarios are consistent with other experimental constraints from dark matter direct detection (DD) and indirect detection (ID) experiments, collider experiments, measure of anomalous magnetic moments and neutrino experiments. We also consider future probes of the scenarios which survive all constraints.

In Section \ref{sec:Models} we present the set of simplified leptophilic models we investigate. Then in Section \ref{sec:constraints} we discuss the constraints on these models that we will consider.  In Section \ref{sec:results} we present the results of our analysis and finally in Section \ref{sec:conclusions} we give our conclusions. 

\section{Models}
\label{sec:Models}

\begin{table}[]
\centering
\begin{tabular}{cccc}
\toprule
DM  particle & Mediator coupling to DM & \multicolumn{2}{c}{Mediator coupling to SM leptons}\\
\cmidrule(r){1-2}\cmidrule(r){2-4}
& & Scalar, $\ell\bar{\ell} \Phi$ & Pseudoscalar, $\ell\gamma^5\bar{\ell} \Phi$\\ 
\midrule
Real scalar & Scalar, $\chi_r\chi_r \Phi$ &  \cellcolor{interesting} DM DD & \cellcolor{interesting} LEP\\ 
Complex scalar & Scalar, $\chi_c^*\chi_c \Phi$ & \cellcolor{interesting} DM DD & \cellcolor{interesting} LEP \\
Dirac fermion & Scalar, $\chi_d\chi_d \Phi$  & \cellcolor{suppressed} $\sigma v\sim v^2$ & \cellcolor{suppressed} $\sigma v\sim v^2$\\ 
Dirac fermion & Pseudoscalar, $\chi_d\gamma^5\chi_d \Phi$ & \cellcolor{interesting} LEP & \cellcolor{interesting} LEP\\
Majorana fermion & Scalar, $\chi_m\chi_m \Phi$ & \cellcolor{suppressed} $\sigma v\sim v^2$ & \cellcolor{suppressed} $\sigma v\sim v^2$\\ 
Majorana fermion & Pseudoscalar, $\chi_m\gamma^5\chi_m \Phi$ & \cellcolor{interesting} LEP & \cellcolor{interesting} LEP\\
\cmidrule(r){1-2}\cmidrule(r){2-4}
& & Vector, $\bar{\ell}\gamma_{\mu}\ell Z^{\prime\mu}$  & Axial-vector, $\bar{\ell}\gamma_{\mu}\gamma^5\ell Z^{\prime\mu}$\\
\midrule
Complex scalar &Vector, $\chi_c^*\partial_\mu\chi_c Z^{\prime\mu}$ & \cellcolor{suppressed} $\sigma v\sim v^2$ & \cellcolor{suppressed} $\sigma v\sim v^2$\\
Dirac fermion &Vector, $\chi_d\gamma_{\mu}\chi_d Z^{\prime\mu}$ & \cellcolor{interesting} DM DD & \cellcolor{interesting} LEP\\
Dirac fermion &Axial-vector, $\chi_d\gamma_{\mu}\gamma^5\chi_d Z^{\prime\mu}$ & \cellcolor{suppressed} $\sigma v\sim v^2$ & \cellcolor{suppressed} $\sigma v\sim m_\ell^2$\\
Majorana fermion & Axial-vector, $\chi_m\gamma_{\mu}\gamma^5\chi_m Z^{\prime\mu}$ & \cellcolor{suppressed} $\sigma v\sim v^2$ & \cellcolor{suppressed} $\sigma v\sim m_\ell^2$\\
\bottomrule
\end{tabular}
\caption{Combinations of possible (spin 0 or 1) mediator couplings to SM leptons (columns) and DM (rows). For each possibility, we indicate the strictest constraint on the parameter space explaining the DAMPE excess (colored green) or indicate 
suppression that rules it out as an explanation of the DAMPE excess (colored red), where DM DD indicates the spin-independent DM direct detection constraints currently provided by PandaX.}
\label{tab:sum}
\end{table}

We model DM by a single species of WIMP which we assume is responsible for the DAMPE signal and the DM abundance in our Universe. We assume that it interacts with Standard Model (SM) leptons by a new massive mediator. Instead of specifying UV-complete models, we study DM models that could explain DAMPE's result in a model-independent way by coupling scalar and fermionic DM particles, $\chi$, to a massive scalar or vector mediator. We require that the couplings between the mediator and the DM, and the mediator and SM leptons satisfy only Hermiticity and Lorentz invariance, and that it is renormalizable. 

The scalar mediator, $\Phi$, couples to the SM leptons by a scalar or pseudoscalar interaction,
\begin{equation}
\mathcal{L}_\text{scalar--SM} = \sum_{\ell = e, \mu, \tau} \bar{\ell} \left(g_{\ell}^s+ig_{\ell}^{p}\gamma^5\right) \ell \Phi.
\end{equation}
There are no Yukawa couplings between the scalar and (only left-handed) neutrinos. In the case of e.g., right-handed neutrinos and a see-saw mechanism, Yukawa interactions between light mass eigenstates and the spin 0 mediator would be suppressed by the see-saw scale or equivalently lightness of neutrino masses.  We pick universal couplings between the mediator and the SM leptons, $g_\ell = g_e = g_\mu = g_\tau$. The scalar couples to DM by either
\begin{equation}\label{eq:la_y0}
\mathcal{L}_\text{scalar--DM} = 
\begin{cases} 
\frac12 \kappa_r \chi_r^2 \Phi + \frac12 \lambda_r \chi_r^2 \Phi^2 & \text{Real scalar}\\
\kappa_c |\chi_c|^2 \Phi + \lambda_c |\chi_c|^2 \Phi^2  & \text{Complex scalar}\\
\bar{\chi}_d \left(g^s_{\chi_d}+ig^p_{\chi_d}\gamma_5\right)\chi_d \Phi & \text{Dirac fermion}\\ 
\frac{1}{2}\bar{\chi}_m \left(g^s_{\chi_m}+ig^p_{\chi_m}\gamma_5\right)\chi_m \Phi & \text{Majorana fermion}\\
\end{cases} 
\end{equation}
for DM that is real scalar $\chi_r$, complex scalar $\chi_c$, Dirac fermion $\chi_d$ or Majorana fermion $\chi_m$, respectively. In the case of scalar DM with a scalar mediator, we define the dimensionless couplings $g_{\chi_{r,c}} \equiv \kappa_{r,c}/m_{\chi}$ so that we may compare it with the dimemsionless coupling to SM leptons. The vector mediator, $Z^\prime$, couples to the SM leptons and left-handed neutrinos by a vector or axial-vector interaction,
\begin{equation}\label{eq:la_lep}
\mathcal{L}_\text{vector--SM} = \sum_{\ell = e, \mu, \tau}\bar{\ell}\gamma_{\mu}\left(g_\ell^{v}+g_\ell^{a}\gamma_5\right)\ell Z^{\prime\mu} + \sum_{\nu = \nu_e, \nu_\mu, \nu_\tau} g_\nu \bar{\nu}\gamma_{\mu}P_L\nu Z^{\prime\mu}.
\end{equation}
As in the case of a scalar mediator, we assume universal lepton couplings. We furthermore assume universal neutrino couplings $g_\nu = g_{\nu_e} = g_{\mu} = g_{\nu_\tau}$. The vector couples to DM by either
\begin{equation}\label{eq:la_y1}
\mathcal{L}_\text{vector--DM} = 
\begin{cases} 
i g^v_{\chi_c}(\chi_c^*\partial_\mu\chi_c - \chi_c\partial_\mu\chi_c^*) Z^{\prime\mu} + 
(g^v_{\chi_c})^2 |\chi_c|^2 Z^{\prime 2} & \text{Complex scalar}\\
\bar{\chi_d}\gamma_{\mu}\left(g^v_{\chi_d}+g^a_{\chi_d}\gamma_5\right)\chi_d Z^{\prime\mu} & \text{Dirac fermion}\\
\frac{1}{2}\bar{\chi_m}g^a_{\chi_m}\gamma_{\mu}\gamma_5\chi_m Z^{\prime\mu} & \text{Majorana fermion}\\
\end{cases} 
\end{equation}
We assume the scalar is complex, so it can be charged under the gauge symmetry. A vector interaction vanishes for a Majorana fermion as the operator is odd under charge symmetry. Our notation for the mediator couplings is that a superscript $s$ denotes a scalar interaction, $p$ denotes pseudoscalar, $v$ denotes vector and $a$ denotes axial-vector. We implemented the models defined in Eq.~\ref{eq:la_y0} and \ref{eq:la_y1} in \texttt{micrOMEGAs-\allowbreak{}4.3.5}\cite{Belanger:2001fz,Barducci:2016pcb} via  \texttt{FeynRules-2.3.27}\cite{Christensen:2008py,Christensen:2009jx} and \texttt{calcHEP}\cite{Belyaev:2012qa}. 

For simplicity, we initially make the following assumptions: 
\begin{enumerate}
\item \label{assumption1} In each scenario, we assume there is a single species of DM and a single mediator.
\item \label{assumption2} We assume that the interactions between the scalar (vector) mediator and leptons or DM is either completely scalar (vector) or completely pseudoscalar (axial-vector), but not a mixture. 
\item \label{assumption3} As discussed, we assume lepton flavor universal couplings between the mediator and SM leptons.
\item \label{assumption4} We assume that $g_{\chi} = g_{\ell}$.
\item \label{assumption5} In keeping with assumption \ref{assumption2} we set $g_\nu = 0$ rather than considering and the lepton interaction simultaneously.
\item We assume no tree-level mixing between the $Z^\prime$ and the $Z$-boson and no tree-level mixing between the SM Higgs boson and our spin 0 mediator. In principle, we could tune tree-level couplings to cancel loop induced mixing that results in DM scattering with quarks.
\end{enumerate}

\noindent The resulting allowed combinations from assumptions \ref{assumption1} and \ref{assumption2} are listed in Table~\ref{tab:sum}, along with an indication of the result of our analysis which are presented in detail in section  \ref{sec:results}.

After this we relax a couple of these assumptions.  First we relax assumption \ref{assumption4}.  In most cases, the relic density $\Omega h^2$ and annihilation cross section $\langle\sigma v\rangle$ depend on the product of mediator couplings to DM and leptons, $g_{\chi} g_{\ell}$, and therefore this assumption has no affect. However, if the mediator is lighter than the DM, DM can annihilate into on-shell pairs of mediators. This process mainly depends on the mediator coupling with DM, $g_{\chi}$. Furthermore constraints from LEP and $\Delta a_{\mu}$, depend on only $g_{\ell}$.  Where relevant we discuss the impact of this assumption and also present results that demonstrate the impact of this assumption when it is most significant.  Secondly while assumption \ref{assumption5} is useful to give an indication of the impact of the lepton interaction alone, it is difficult to generate such scenario in UV complete model, due to SU(2) gauge invariance. Therefore we subsequently break this condition and study scenarios with both lepton and neutrino interactions. 

\section{Constraints}
\label{sec:constraints}

We wish to find points in the parameter spaces of our DM models that satisfy existing experimental constraints from DM searches, predict the correct relic abundance of DM and explain the DAMPE excess. We detail the relevant constraints in the following subsections.

\subsection{Dark matter abundance}

We require that the thermal relic density of DM, calculated in \texttt{micrOMEGAs-\allowbreak{}4.3.5}, satisfies $\Omega h^2=0.1199 \pm 0.0022$ in accordance with measurements from Planck\cite{Ade:2015xua}. This translates into a strict relation between the mediator mass and couplings.  
  
\subsection{DAMPE excess}\label{sec:dampe}

To accommodate the amplitude of the DAMPE excess, we assume a local subhalo with an enhanced density of DM. The required DM density is about $17$ -- $35$ times greater than the local density of DM for thermally produced DM with $\sigmav \simeq 3 \times 10^{-26}\cmCubePerSec$ --- this results in the required boost factor\cite{Yuan:2017ysv}. This assumes that the DM annihilation cross section is not suppressed in the low-velocity limit. Thus, we require $\sigmav_{v\to0} \gtrsim \sigmav \approx [2,4]\times 10^{-26}\cmCubePerSec$ --- the annihilation cross section at freeze-out required for the correct relic density. The necessary cross sections are doubled in the case that the DM is not self-conjugate as the flux would reduce by a factor of two. 

For points that satisfy the DM relic density constraint, we calculate the annihilation cross section $\sigmav$ in \texttt{micrOMEGAs-\allowbreak{}4.3.5}. 
Following\cite{Berlin:2014tja}, of the 20 combinations of operators that link DM with leptons in Table~\ref{tab:sum}, there are 8 in which the low-velocity annihilation cross section $\sigmav$ is suppressed by the velocity, and 2 in which it is suppressed by helicity (see e.g.~Table IV of\cite{Kumar:2013iva}). In Table~\ref{tab:sum}, we denote models suppressed by velocity by $\sigma v \sim v^2$ and models suppressed by helicity by $\sigma v \sim m_\ell^2$.

\subsection{Dark matter direct detection}

In DM direct detection experiments, leptophilic DM may predominantly scatter with bound electrons in the target. The recoiling electrons are either ejected from the target atoms or remain bound if the recoil is taken by the atom as a whole. In the former case, null results from XENON100 constrain the scattering cross section to be $\sigma^0_{\chi e}\lesssim 10^{-34}\,\text{cm}^2$\cite{Aprile:2015ade} at the $90\%$ confidence level. Of the interactions we consider, XENON100 is most sensitive\cite{Kopp:2009et} to the axial-vector coupling (Eq.~\ref{eq:la_y0} and \ref{eq:la_y1}). The axial-vector coupling predicts that
\begin{equation}\label{eq:dd_chie}
\sigma_{\chi e}= 3 g_{\chi}^a g_\ell^a \frac{m_e^2}{\pi M_{Z^\prime}^4} \approx 3 g_{\chi}^a g_\ell^a \left(\frac{M_{Z^\prime}}{10\gev}\right)^{-4} \times 3.1\times10^{-39}\,\text{cm}^2.
\end{equation}
Thus, even for a mediator mass of $10\gev$ and couplings $g_{\chi}^a g_\ell^a = 1$, the cross section would be far lower than the XENON100\cite{Aprile:2015ade} limit.

Leptophilic DM can, however, scatter with quarks in DD experiments through lepton loops. As discussed in\cite{Kopp:2009et}, from the models satisfying our low-velocity annihilation cross section requirement, only one model has a 1-loop cross section for scattering of DM on a nucleus, and four models have 2-loop cross sections. 

We calculate the loop induced $\chi n \to \chi n $ cross section $\sigma_{\chi n}$ using expressions given in\cite{Kopp:2009et}\footnote{The expressions are obtained by assuming the nuclear structure form factors $F(q)=1$ and $\tilde{F}(q)=1$ and vector coefficients, for which there is no uncertainty. We have checked that with the kinetic recoil energy of the nucleus $E_d\in[1,10]$ keV for Xenon, $F(q)$ lies in $[0.9995,0.9950]$ using formula in\cite{ENGEL1991114} and $\tilde{F}(q)$ lies in $[0.9999,0.9998]$ using formula in\cite{Frandsen:2012db}.},
\begin{align}\label{eq:dd_chin}
\sigma_{\chi n} =&{}\frac{\alpha_{\rm em}^2 Z^2 \mu_{N}^2}{\pi^3 A^2 M^4} \sum_{\ell=e,\mu,\tau}\nonumber\\
&
\begin{cases} 
\frac{1}{9} \left( g_{\chi}^vg_{\ell}^v \log \frac{m_\ell^2}{\mu^2}\right)^2  & \text{mediator:} \, Z^\prime,\;\text{DM:} \, \chi_d\\
\left(\frac{\pi\alpha_{\rm em}Z\mu_N v}{6\sqrt{2}}\right)^2 \left[\left(\frac{g_{\chi_{d,m}}^sg_{\ell}^s}{m_{\ell}}\right)^2  +\frac{2}{3}\left(\frac{g_{\chi_{d,m}}^p g_{\ell}^s\mu_N v}{m_{\ell}m_{\chi}}\right)^2    \right]            & \text{mediator:}\,\Phi,\;\text{DM:}\, \chi_{d,m}\\
\frac{1}{4}\left(\frac{\pi\alpha_{\rm em}Z\mu_N v}{6\sqrt{2}}\right)^2  \left(\frac{\kappa_{r,c} g_{\ell}^s}{m_{\ell}M}\right)^2 & \text{mediator:}\,\Phi,\;\text{DM:} \, \chi_{r,c}\\
\end{cases}
\end{align}
 where $\alpha_{\rm em}$ is the fine structure constant, $M$ is the mediator mass, $\mu_N \equiv m_N m_{\chi} / (m_N+m_{\chi})$ is the reduced mass of the DM-nucleus two particle system,  $v=0.001c$ is the velocity of the DM near the Earth, $\mu$ is the scale at which the logarithmic divergence is cut off, $m_N$, $Z$ and $A$ are the target nucleus' mass, charge and mass number respectively. The most stringent direct detection constraints on DM with mass larger than 10 GeV are currently provided by PandaX\cite{Cui:2017nnn,Fu:2016ega}, though XENON1T\cite{Aprile:2017iyp} and LUX\cite{Akerib:2016vxi} established similar limits with related technology. Therefore we take $Z=54$, $A=131$ and $m_{N}=131\gev$ for a target nucleus of Xenon and $\mu=M/\sqrt{g_{\chi}^v g_{\ell}^v}$. We compare the predicted cross sections with 90\% confidence limits on spin independent elastic DM-nucleon cross section. 

The DD constraints on leptophilic DM scattering with quarks are equivalent to constraints upon mixing between the SM $Z$ and Higgs bosons and our spin 1 and spin 0 mediator, respectively. The DD constraints are, however, stronger than precision measurements\cite{Hook:2010tw}. As noted in the introduction, though, DD constraints could be evaded by fine-tuning the tree-level mixing against loop-induced corrections.

\subsection{Dark matter indirect detection}
As well as the potential signal from DAMPE, there are constraints on the annihilation of DM from other ID experiments.  In our scenarios, DM may annihilate to leptons.  Hadronic tau decays result in gamma-rays from pion decay.  Amongst others H.E.S.S.\cite{Abdallah:2016ygi}, Fermi-LAT\cite{Ackermann:2015zua} and IceCube\cite{Aartsen:2017ulx} searched for photons and neutrinos from regions in which DM may be abundant, such as the Galactic Centre, dwarf Spheroidal galaxies, and the Sun.

An excess of gamma rays was reported by the Large Area Telescope (LAT) on board the Fermi Gamma Ray Space mission\cite{Goodenough:2009gk, Hooper:2010mq, Boyarsky:2010dr, Abazajian:2012pn, Hooper:2013rwa, Gordon:2013vta, Daylan:2014rsa, Calore:2014xka}.  The origin of this relatively low energy (2-3 GeV) gamma ray flux was thought to be the Galactic Center.  Their source is under extensive debate in the literature\cite{Fornasa:2016ohl, Gomez-Vargas:2013bea, Abazajian:2014fta, Ipek:2014gua, Ko:2014gha, Cline:2014dwa, Ko:2014loa, arXiv:1405.0272, Kong:2014haa, Boehm:2014bia, Ghosh:2014pwa, Wang:2014elb, Fields:2014pia, Berlin:2014tja, Carlson:2014cwa, Petrovic:2014uda, arXiv:1312.7488, arXiv:1402.4500, arXiv:1402.6671, arXiv:1403.1987, arXiv:1403.3401, arXiv:1403.5027, arXiv:1404.1373, arXiv:1404.2018, arXiv:1404.2067, arXiv:1404.2318, arXiv:1404.2572, arXiv:1404.3362, arXiv:1404.5503, arXiv:1404.6528, arXiv:1405.1030, arXiv:1405.1031, arXiv:1405.5204, arXiv:1405.4877, arXiv:1405.6240, arXiv:1405.6709, arXiv:1405.7059, arXiv:1405.7370, arXiv:1406.0507, arXiv:1406.1181, arXiv:1406.2276, arXiv:1406.4683, arXiv:1406.5662, arXiv:1406.6372, arXiv:1406.6408, Zhou:2014lva, arXiv:1408.4929, arXiv:1408.5795, arXiv:1409.1406, arXiv:1409.7864, arXiv:1410.3239, arXiv:1410.4842, arXiv:1411.2592, arXiv:1411.2619, arXiv:1411.4647, arXiv:1412.1663, arXiv:1412.5174, arXiv:1501.00206, arXiv:1501.07275, arXiv:1501.07413, arXiv:1502.05682, arXiv:1502.05703, arXiv:1503.06348, arXiv:1503.08213, arXiv:1503.08220, arXiv:1504.03610, arXiv:1504.03908, arXiv:1505.04620, arXiv:1505.04988, arXiv:1505.07826, arXiv:1506.05119, Cao:2015loa, arXiv:1507.01793, arXiv:1507.02288, arXiv:1507.04644, arXiv:1507.07008, arXiv:1507.07922, arXiv:1507.08295, arXiv:1507.05616, arXiv:1508.05716, Dutta:2015ysa, arXiv:1509.02928, arXiv:1509.05076, arXiv:1509.09050, arXiv:1510.00714, arXiv:1510.04698, arXiv:1510.06424, arXiv:1510.07562, arXiv:1511.02938, arXiv:1511.09247, Mora:2015vhq, arXiv:1512.01846, arXiv:1512.02899, arXiv:1512.04966, arXiv:1512.06825, arXiv:1601.05089, arXiv:1601.05797, arXiv:1602.00590, arXiv:1602.04788, arXiv:1602.05192, arXiv:1603.08228, arXiv:1604.00744, arXiv:1604.01039, arXiv:1604.01026, arXiv:1604.01402, arXiv:1604.04589, arXiv:1604.06566, arXiv:1606.09250, arXiv:1607.00737, arXiv:1608.00786, arXiv:1608.07289,Karwin:2016tsw,Cirelli:2014lwa}.  The most recent Fermi Collaboration paper, however, claims that the excess emission is also present in ``control regions along the Galactic plane, where a dark-matter signal is not expected''\cite{TheFermi-LAT:2017vmf}.
Assuming that the above excess is explained by standard astrophysical sources, the most stringent limits on dark matter annihilation into a pair of gamma rays comes from Fermi-LAT observations of dwarf spheroidal satellite galaxies (dSphs) of the Milky Way.  This is because it is assumed that dark matter dominated these dwarf galaxies.  The Fermi-LAT upper limit on the dark matter annihilation cross section combines the analysis of 15 Milky Way dSphs\cite{Ackermann:2015zua}.  The latter provides the most stringent constraint for dark matter annihilating into $\tau$ leptons (or quarks).  
Additional, comparably strong limits come from AMS-02 and the cosmic microwave background (CMB) and\cite{Elor:2015bho}.  All these limits, however, are fairly weak for a DM mass of a TeV.

Observation of charged cosmic leptons hinted anomalies for more than a decade.  The electron and positron fluxes have been especially controversial for some time\cite{Golden:1992zm, Alcaraz:2000bf, Boezio:2001ac, Grimani:2002yz, Barwick:1997ig, Beatty:2004cy, Adriani:2008zr, Delahaye:2008ua, Delahaye:2010ji, Mertsch:2010qf, Timur:2011vv, Aguilar:2002ad, Torii:2008xu, Aharonian:2008aa, Aharonian:2009ah, Ackermann:2010ij, Accardo:2014lma, Aguilar:2014mma, Aguilar:2014fea, Aguilar:2015ooa,Cirelli:2008pk}.  The growth of the positron-to-electron fraction and the increase of the positron spectral index above 100 GeV in the Fermi-LAT data are both considered as signs of unexplained sources\cite{Accardo:2014lma, Aguilar:2014mma}.  The nature of these new cosmic ray sources is still a subject of debate.  They may be standard astrophysical sources (supernova remnants and/or pulsars), or sources harboring new physics (dark matter annihilation)\cite{Serpico:2011wg, Belotsky:2014nba, Mambrini:2015sia}.  Our work is motivated by the latter interpretation.

\subsection{Anomalous magnetic moments of leptons}
\label{sec:amu}
Since the mediators couple to leptons, they could have a significant impact on the anomalous magnetic moments of the leptons.
Since we consider a lepton universal coupling, we consider only the magnetic moment of the muon; we do not consider weaker constraints on the anomalous magnetic moments of the electron or tau. The discrepancy between experiment and the SM prediction for the magnetic moment of the muon is about\cite{Bennett:2006fi,Patrignani:2016xqp,Davier:2010nc}
\begin{equation}
\Delta a_\mu = 28.8 \pm 5.4 \pm 3.3 \pm 4.9 \times 10^{-10},
\end{equation}
where the first error is statistical, the second is systematic and the last is the estimated theoretical uncertainty in the SM calculation given in\cite{Davier:2010nc}. Combining these errors in quadrature one obtains a total uncertainty of $8.0 \pm 10^{-10}$, making the deviation between experiment and the SM prediction approximately $3.6\sigma$.  However it is important to note that the SM calculation and the associated uncertainty involve challenging estimates of the hadronic contributions, and different values can be found in the literature\cite{Hagiwara:2011af,Jegerlehner:2011ti,Benayoun:2012wc,Kurz:2014wya}.  While the estimates cited here all show a significant deviation, it is not universally accepted in the wider community that the hadronic uncertainties are under control.  Moreover the SM is not regarded as excluded, so it would seem strange to rule out BSM theories which lie between the SM prediction and the experimentally measured one.

For this reason we consider two constraints.  First we directly use the combined uncertainty given above to form a $2\sigma$ interval within which the BSM model can explain the deviation.  Secondly to make a much more conservative exclusion, which reflects more conservative views about the hadronic uncertainty, we consider a theoretical uncertainty of $25 \times 10^{-10}$ where most of the deviation between the theory prediction and the experimental value is covered by the theory uncertainty.  While this is a very conservative estimate of the theory uncertainty, it still provides a non-trivial constraint on BSM models, allowing one to exclude models.     

The one-loop BSM contribution to the magnetic moment, from diagrams involving the mediator, can be approximated by\cite{Leveille:1977rc,arXiv:1404.1373},
\begin{equation}\label{eq:muongm2}
\Delta a_{\mu} = \left(\frac{m_{\mu}} {2\pi M}\right)^2
\begin{cases}
    \frac{1}{3} (g_\ell^v)^2 - \frac{5}{3} (g_\ell^a)^2 & \text{Vector mediator} \\
    -\left(\frac{7}{12}+ \ln\frac{m_\mu}{M}\right) (g_\ell^s)^2 + \left(\frac{11}{12} - \ln\frac{m_\mu}{M}\right) (g_\ell^p)^2 & \text{Scalar mediator}\\ 
\end{cases}
\end{equation}
for a vector and vector mediator, respectively, where $m_{\mu}$ is the muon mass and $M$ is the mediator mass and we have neglected terms\footnote{The exact expressions can be found in\cite{Leveille:1977rc,Grifols:1982vx}.} of order ${\cal O}(m^3_{\mu}/M^3)$. For a vector mediator, a vector interaction with electrons improves agreement with measurement, whereas an axial-vector one worsens it. For a scalar mediator, a pseudo-scalar interaction with electrons improves agreement, whereas a scalar one worsens it. 

\subsection{Collider searches}

Whilst there were no dedicated searches for leptophilic $Z^\prime$ bosons at LEP, searches in the framework of four-fermion operators\cite{LEP:2003aa} require that\cite{Buckley:2011vc,Freitas:2014jla,Freitas:2014pua} 
\begin{equation}
    g_e^{s,p} \lesssim 
    \begin{cases}
        {2.7\times10^{-4}M_{\Phi}}/{1\gev} & M_{\Phi} \gtrsim 200\gev\\
        {7.3\times10^{-4}M_{\Phi}}/{1\gev}  & 100\gev \lesssim M_{\Phi} \lesssim 200\gev\\
    \end{cases}
\end{equation}
for a scalar mediator, and 
\begin{equation}
    g_e^{v} \lesssim 
    \begin{cases}
        {2.0\times10^{-4}M_{Z^\prime}}/{1\gev} & M_{Z^\prime} \gtrsim 200\gev\\
        {6.9\times10^{-4}M_{Z^\prime}}/{1\gev}  & 100\gev \lesssim M_{Z^\prime} \lesssim 200\gev\\
    \end{cases}
\end{equation}
\begin{equation}
    g_e^{a} \lesssim 
    \begin{cases}
        {2.4\times10^{-4}M_{Z^\prime}}/{1\gev}  & M_{Z^\prime} \gtrsim 200\gev\\
        {6.9\times10^{-4}M_{Z^\prime}}/{1\gev}  & 100\gev \lesssim M_{Z^\prime} \lesssim 200\gev\\
    \end{cases}
\end{equation}
for a vector mediator. The latter case covers $Z^\prime$ bosons lighter than the maximum LEP centre-of-mass energy, $209\gev$. In this regime, the constraint may be much stronger if the $Z^\prime$ boson mass lies close to one of the LEP centre-of-mass energies, which were 130, 136, 161, 172, 183, 189, and 192--209 GeV. LEP monophoton constraints\cite{Fox:2011fx}, on the other hand, are irrelevant since our DM mass is well above the maximum LEP centre-of-mass energy. 

 At a future $e^+e^-$ collider, such as ILC with a centre-of-mass energies up to 1\tev\cite{Behnke:2013xla}, these constraints are expected to increase significantly. The reach of ILC with a luminosity of 500 fb$^{-1}$ are\cite{Freitas:2014jla}
\begin{equation}
    g_e^{s,p} \lesssim 
    \begin{cases}
        {3.4\times10^{-5}M_{\Phi}}/{1\gev} & M_{\Phi} \gtrsim 1\tev\\
        {9.1\times10^{-5}M_{\Phi}}/{1\gev}  & 100\gev \lesssim M_{\Phi} \lesssim 1\tev\\
    \end{cases}
\end{equation}
for a scalar mediator, and 
\begin{equation}
    g_e^{v} \lesssim 
    \begin{cases}
        {2.2\times10^{-5}M_{Z^\prime}}/{1\gev} & M_{Z^\prime} \gtrsim 1\tev\\
        {7.6\times10^{-5}M_{Z^\prime}}/{1\gev}  & 100\gev \lesssim M_{Z^\prime} \lesssim 1\tev\\
    \end{cases}
\end{equation}
\begin{equation}
    g_e^{a} \lesssim 
    \begin{cases}
        {2.7\times10^{-5}M_{Z^\prime}}/{1\gev}  & M_{Z^\prime} \gtrsim 1\tev\\
        {7.6\times10^{-5}M_{Z^\prime}}/{1\gev}  & 100\gev \lesssim M_{Z^\prime} \lesssim 1\tev\\
    \end{cases}
\end{equation}
for a vector mediator.

The LHC could be sensitive to a leptophilic $Z^\prime$ or scalar mediator produced as bremsstrahlung from a lepton produced by Drell-Yann. The mediator could subsequently decay to combinations of lepton pairs and MET. A detailed study\cite{Bell:2014tta}, however, found that the limits were negligible for $M_{Z^\prime} \gtrsim 100\gev$. We assume that the limits for a scalar mediator would be similar (see e.g.,\cite{Aad:2012xsa,Chatrchyan:2012mea}). A leptophilic scalar mediator could potentially modify Higgs branching ratios if $M_\Phi \le m_h / 2$; however, we assume no coupling between the Higgs and mediator at tree-level.

\subsection{Trident production}

Constraints from so-called trident production, $\nu_\mu N \to \nu_\mu \mu \mu N$\cite{Altmannshofer:2014pba}, place severe restrictions on the $Z^\prime$ coupling to muons for $g_\mu$,
\begin{equation}
    g_\mu \lesssim \frac{M_{Z^\prime}}{1\,\text{TeV}}.
\end{equation}
This constraint applies only if the $Z^\prime$ couples to neutrinos.

\section{Results}
\label{sec:results}

\begin{figure}[tbp]
\centering 
\includegraphics[width=.45\textwidth]{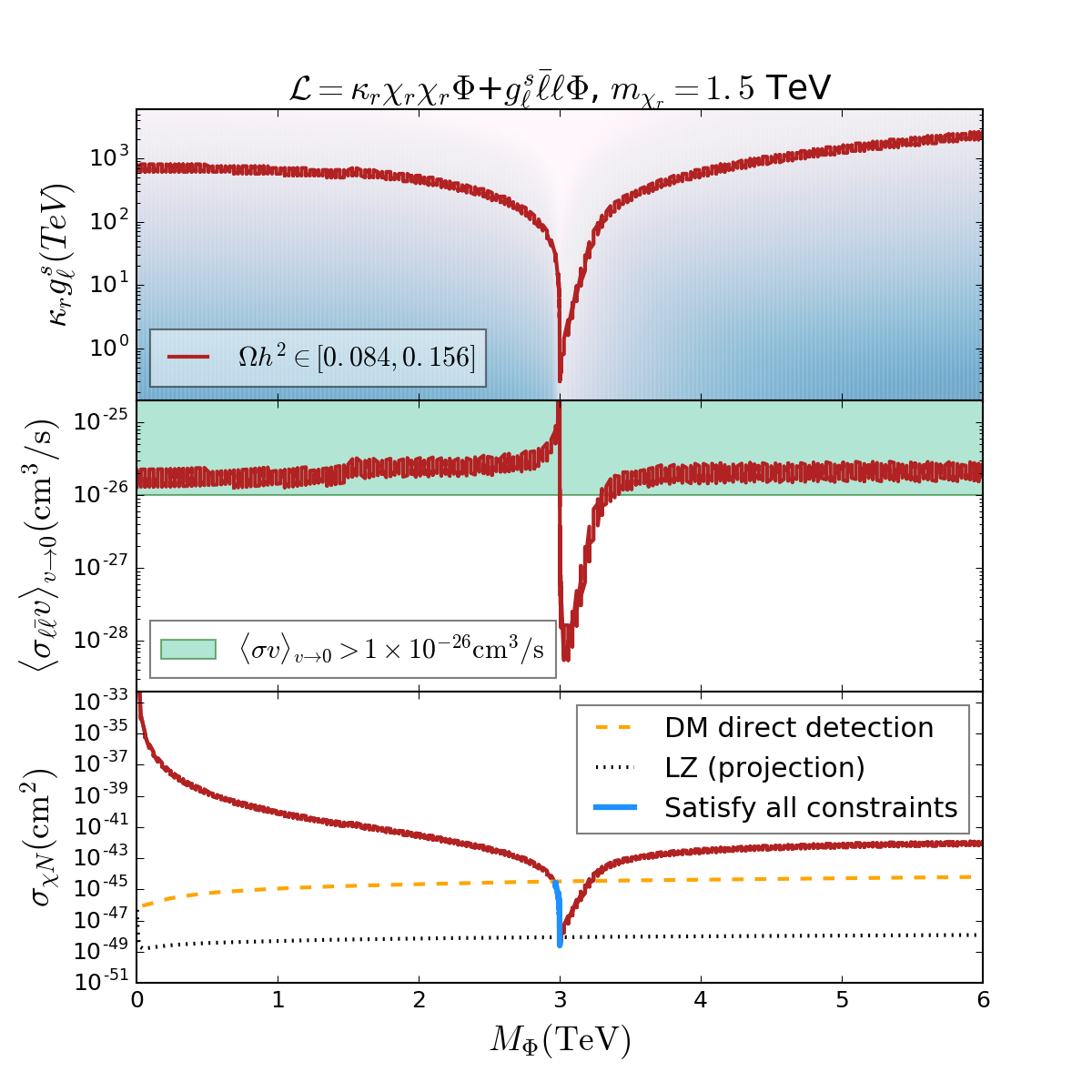}
\hfill
\includegraphics[width=.45\textwidth]{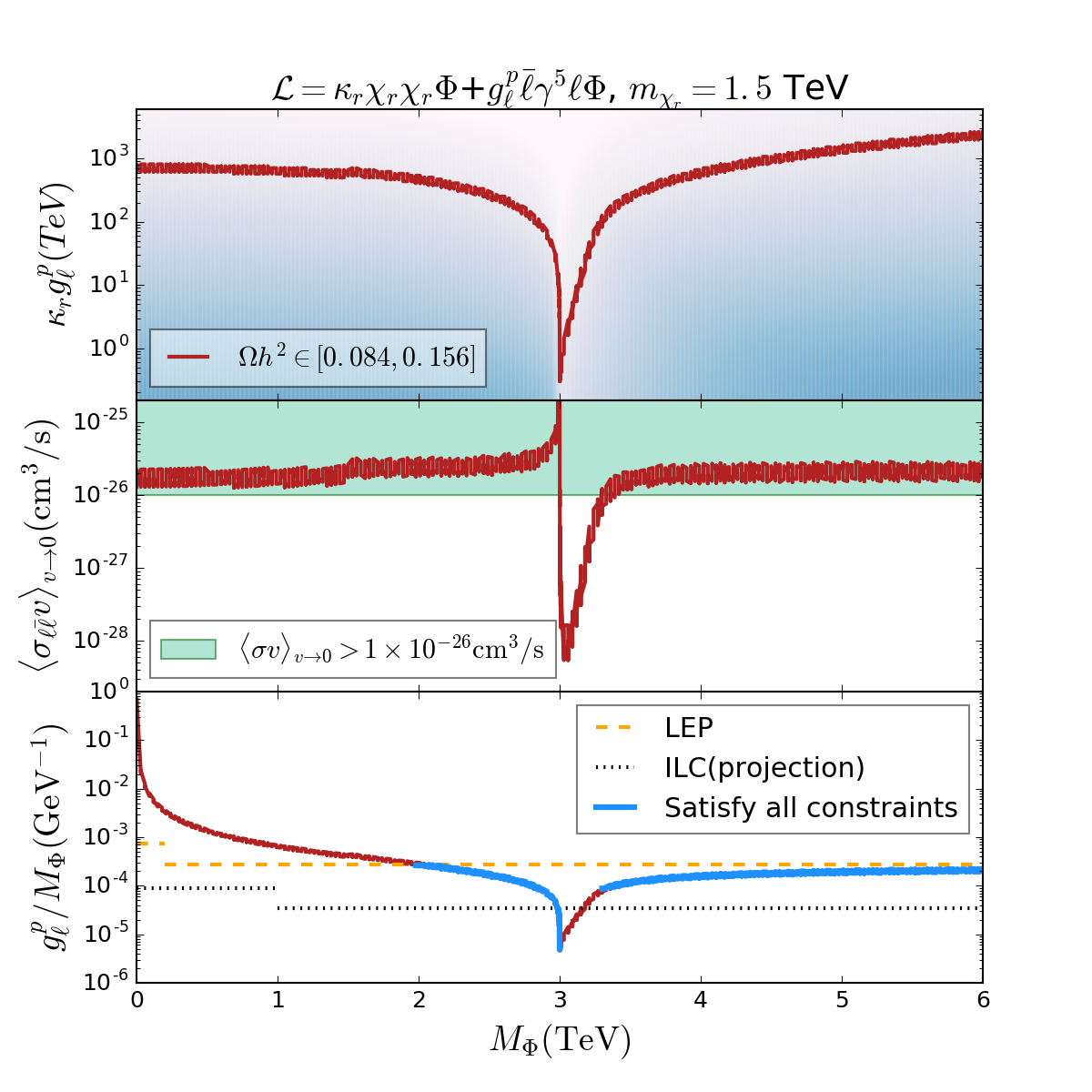}\\
\includegraphics[width=.45\textwidth]{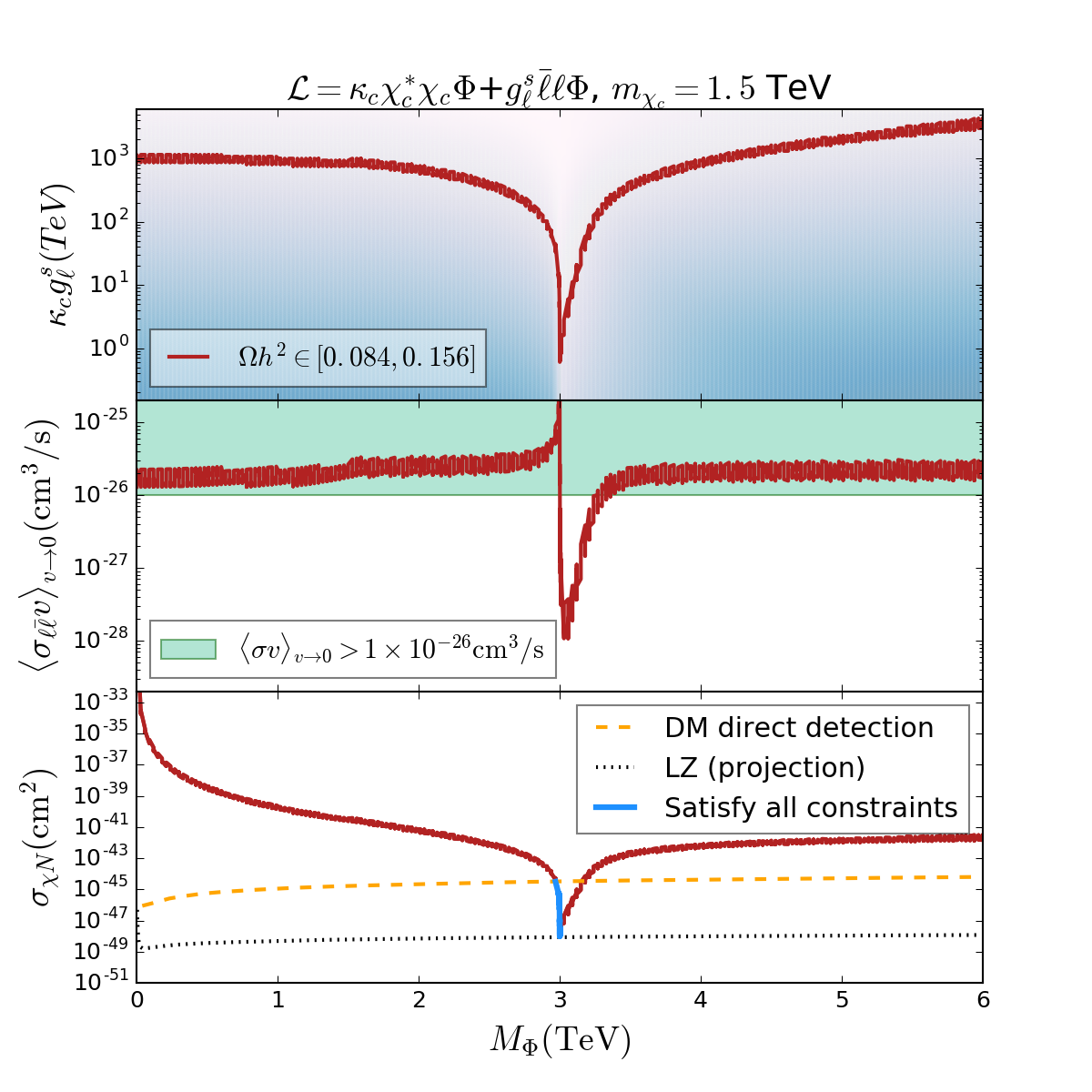}
\hfill
\includegraphics[width=.45\textwidth]{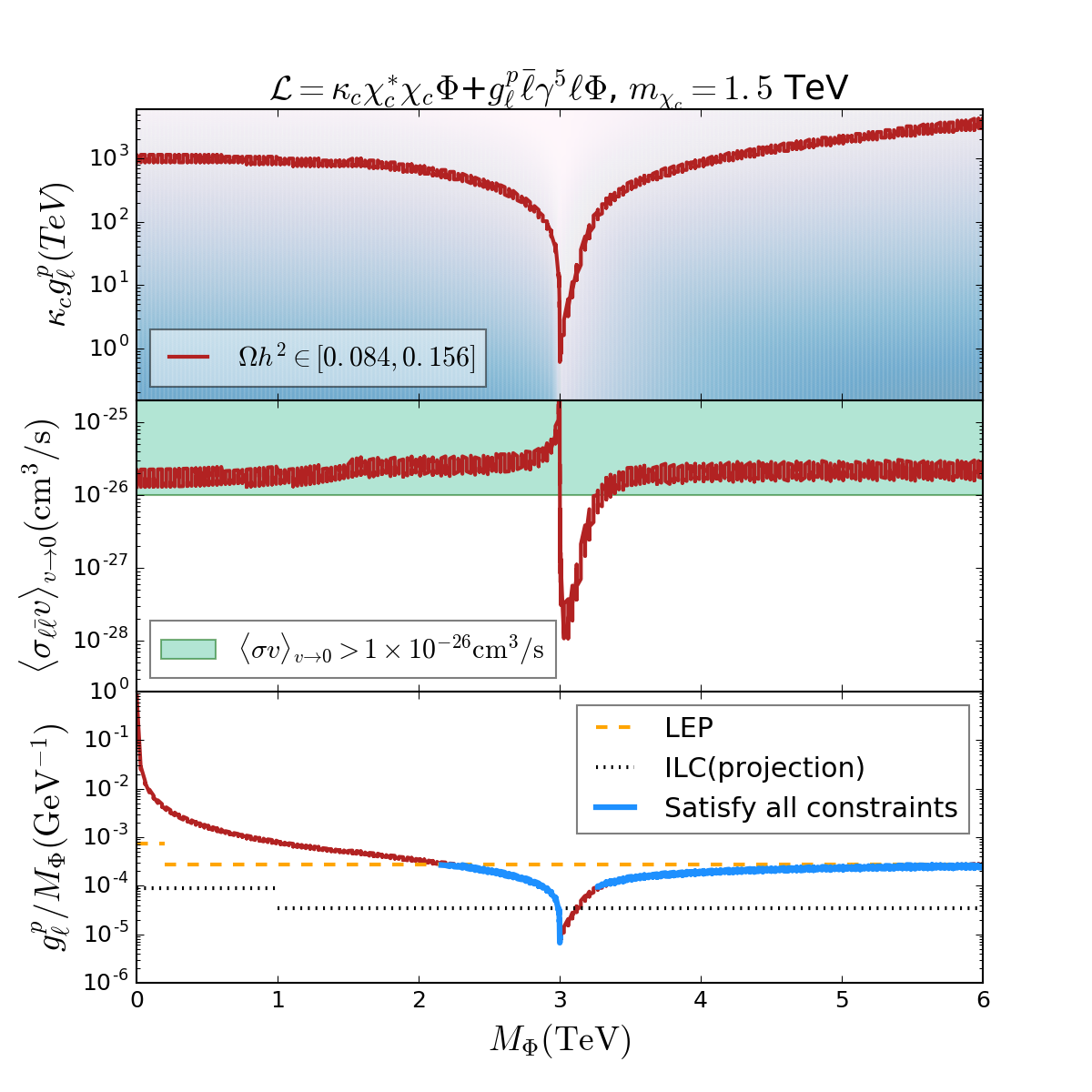}
\caption{\label{fig:spin0_DM0} A scalar mediator scalar coupled to real scalar DM (top) and complex scalar DM (bottom). The  mediator coupling with leptons is scalar (left) or pseudoscalar (right). Each panel shows DM properties as a function of mediator mass. The blue color gradient in the first panel shows relic density, where darker blue means greater relic density.}
\end{figure}

\begin{figure}[tbp]
\centering 
\includegraphics[width=.45\textwidth]{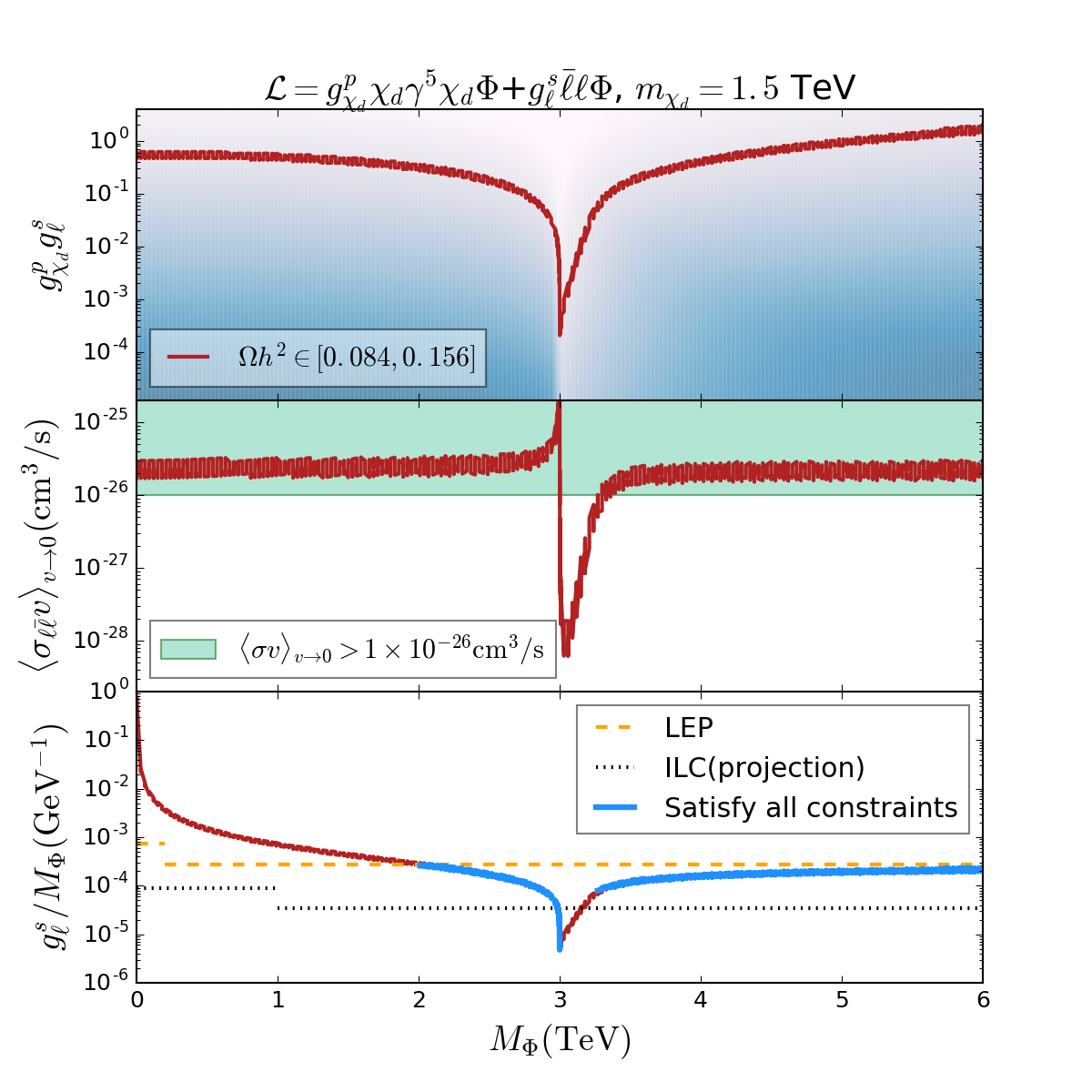}
\hfill
\includegraphics[width=.45\textwidth]{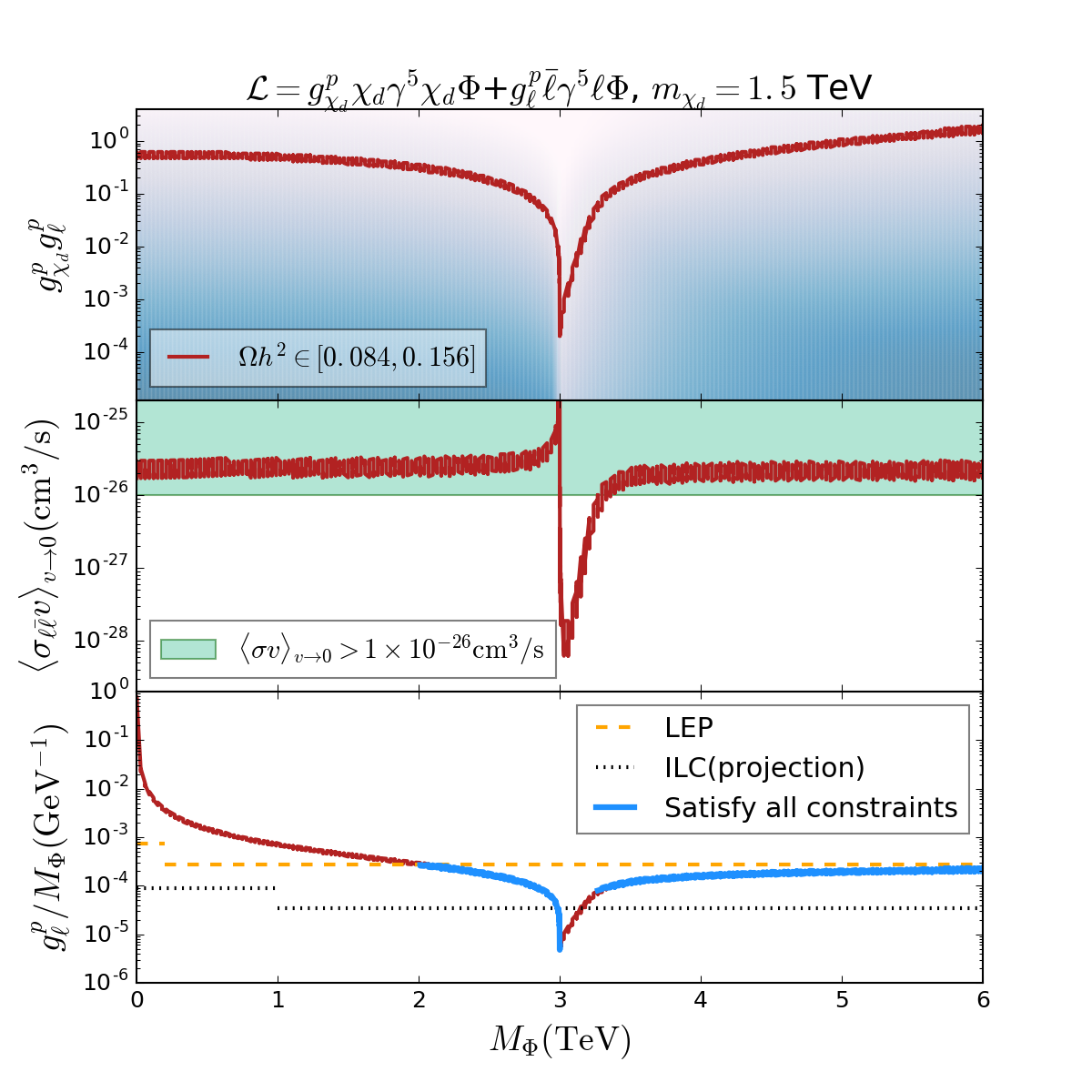}\\
\includegraphics[width=.45\textwidth]{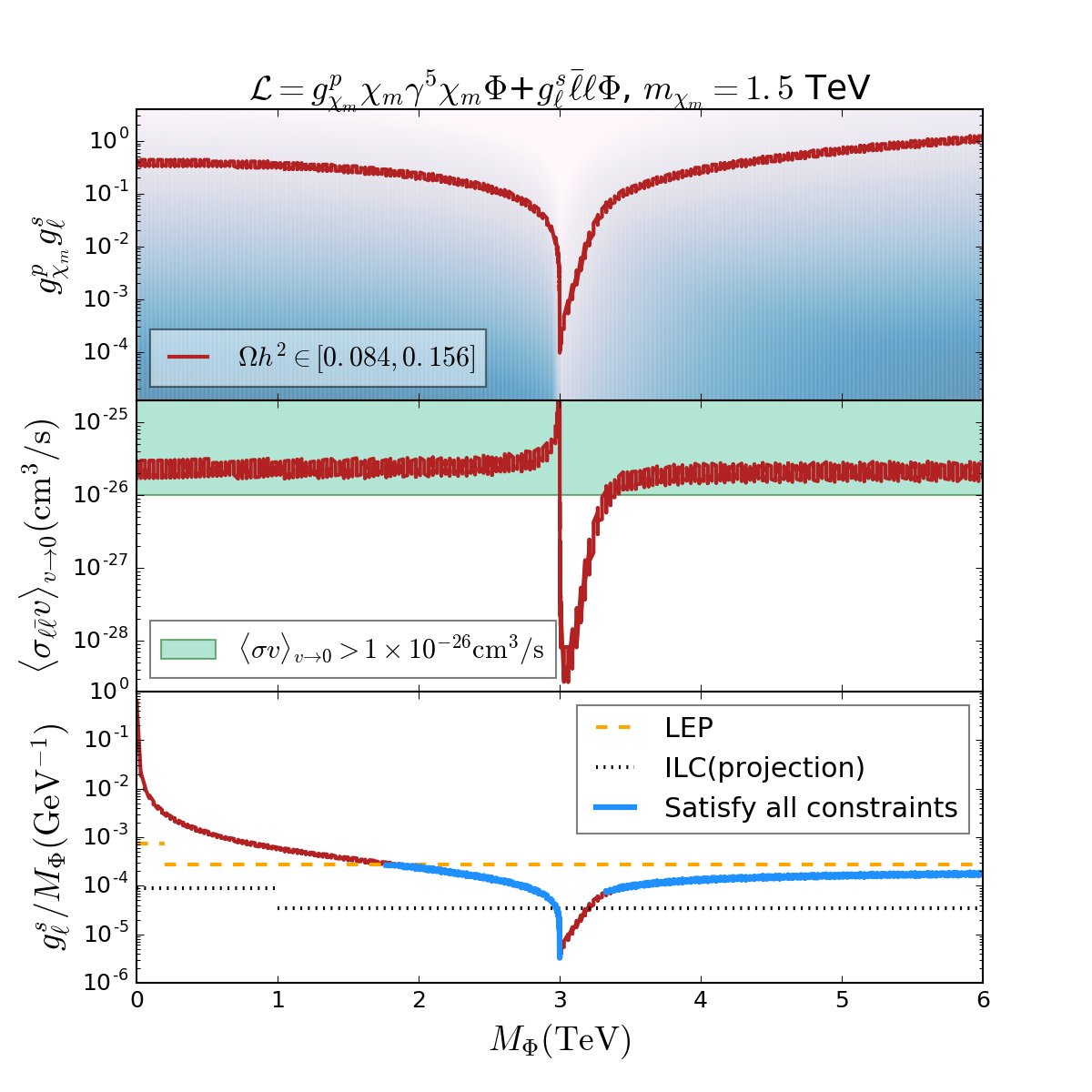}
\hfill
\includegraphics[width=.45\textwidth]{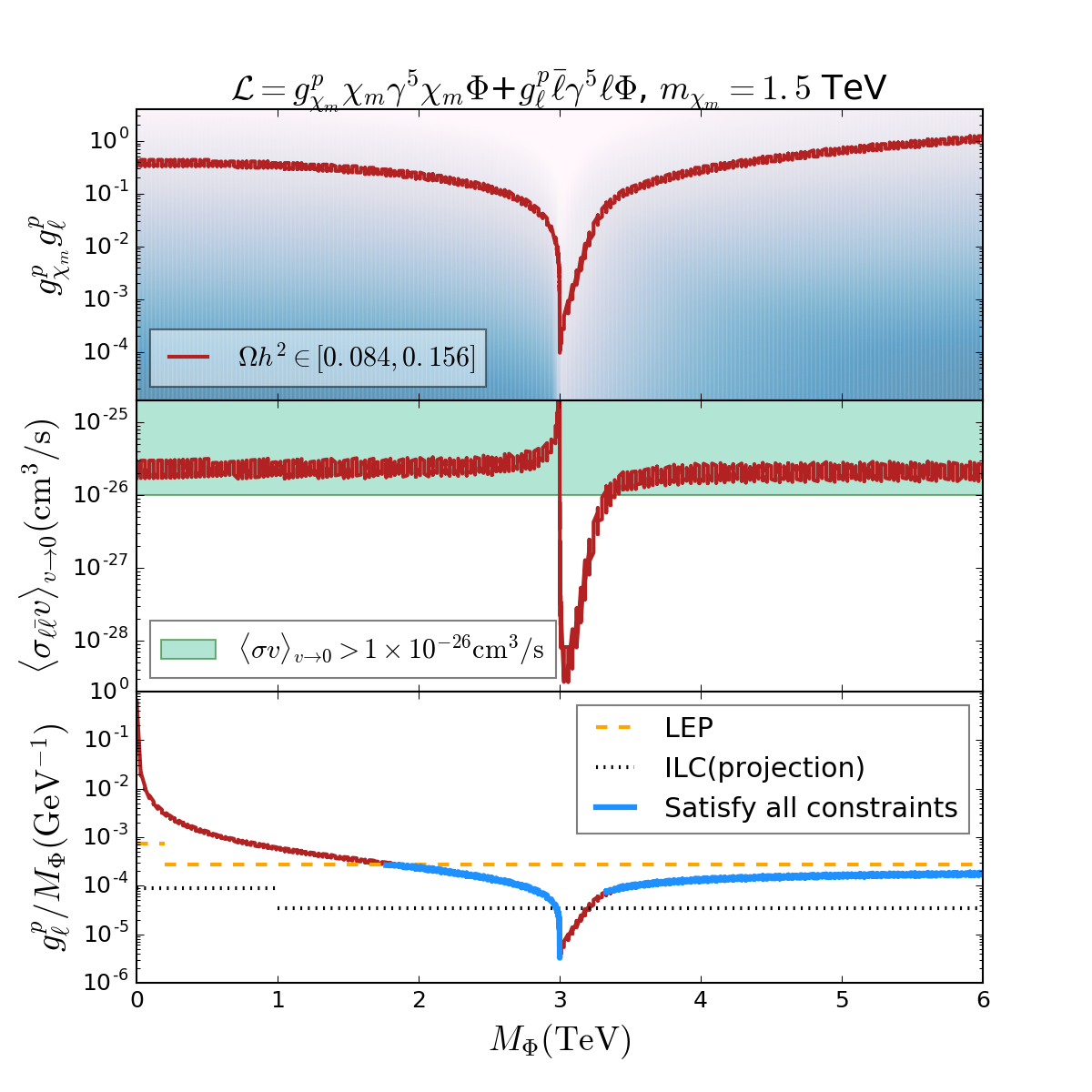}
\caption{\label{fig:spin0_DMfermion} A scalar mediator pseudoscalar coupled to Dirac fermion DM (top) and Majorana fermion DM (bottom). The mediator coupling with leptons is scalar (left) or pseudoscalar (right). Each panel shows DM properties as a function of mediator mass.  The blue color gradient in the first panel shows relic density, where darker blue means greater relic density.}
\end{figure}

\begin{figure}[tbp]
\centering 
\includegraphics[width=.45\textwidth]{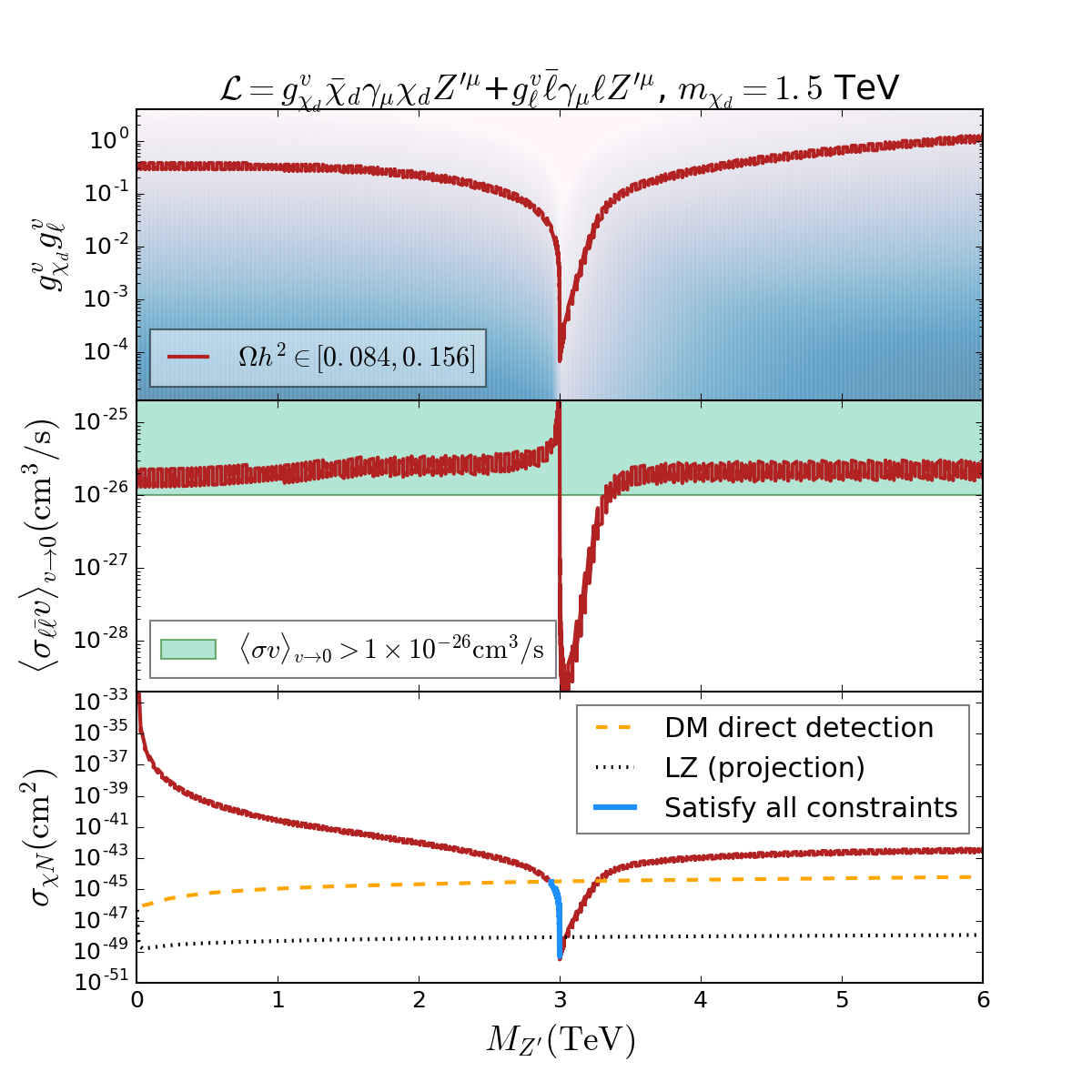}
\hfill
\includegraphics[width=.45\textwidth]{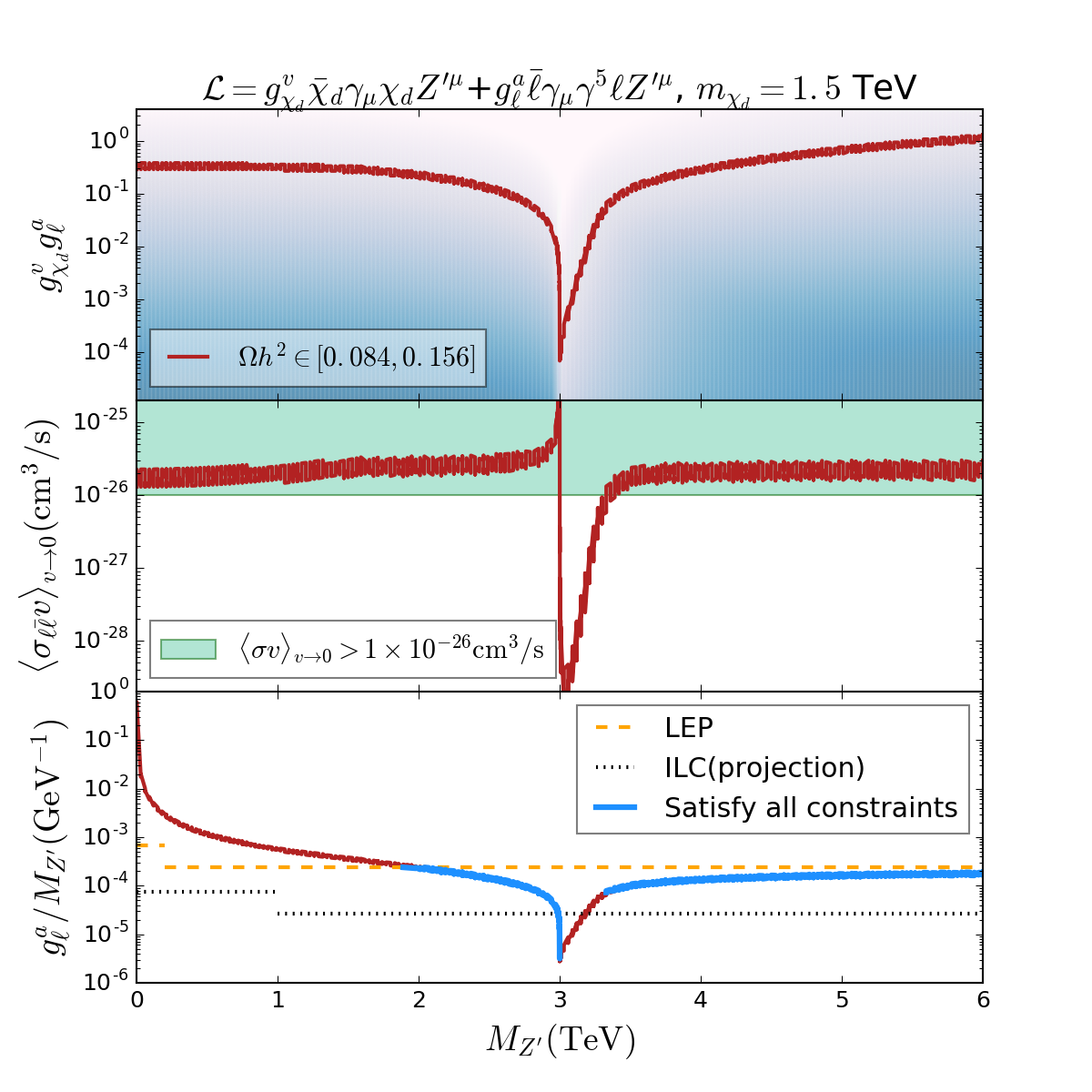}\\
\includegraphics[width=.45\textwidth]{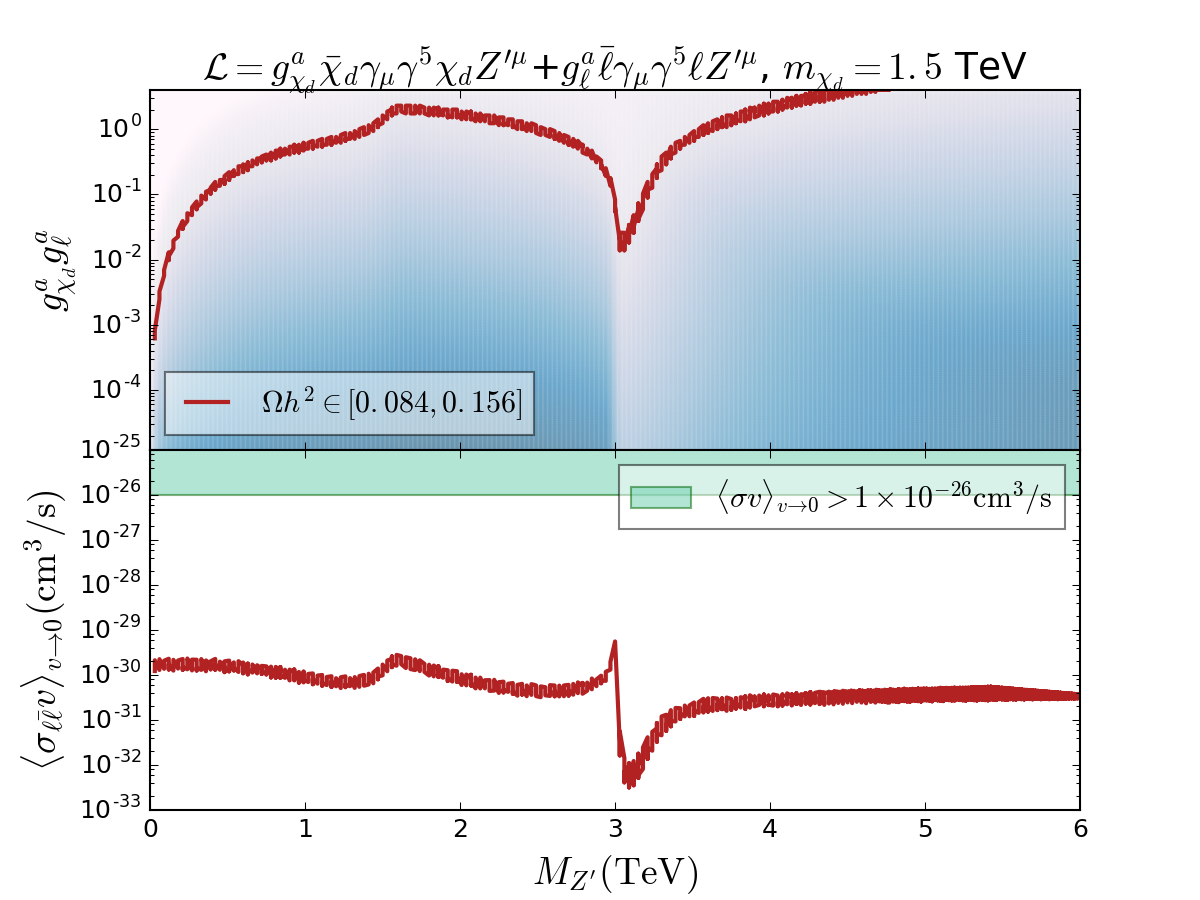}
\hfill
\includegraphics[width=.45\textwidth]{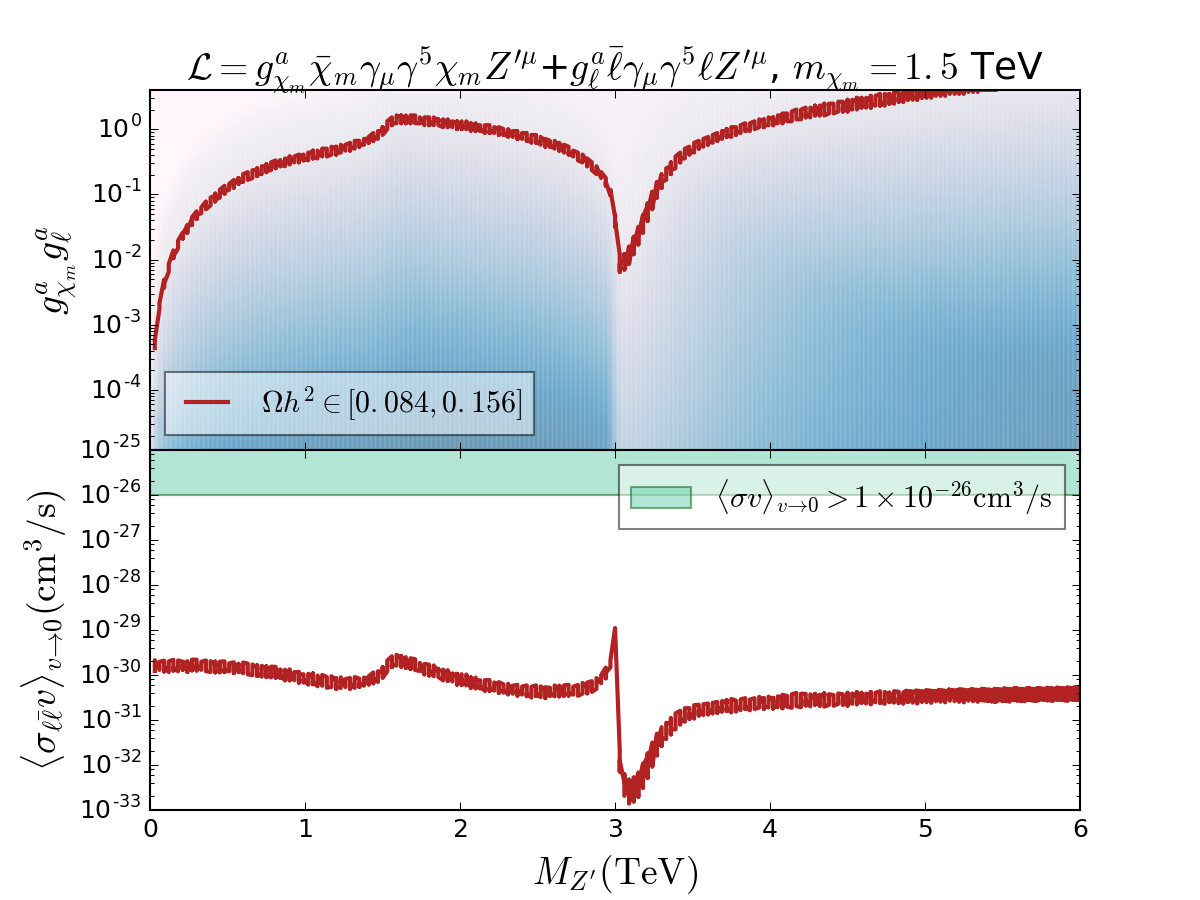}
\caption{\label{fig:spin1_DMfermion} A vector mediator vector coupled to Dirac fermion DM (top), axial-vector to Dirac fermion DM (bottom left) and  Majorana fermion DM (bottom right). The mediator coupling with leptons is vector (top left) or axial-vector (others). Each panel shows DM properties as a function of mediator mass.  The blue color gradient in the first panel shows relic density, where darker blue means greater relic density.}

\end{figure}
The scan of the DM models was performed with the \texttt{EasyScan\_HEP} setup used in\cite{Han:2016gvr}. In Fig.~\ref{fig:spin0_DM0}, we show properties of a real (top) and complex (bottom) scalar DM particle interacting with SM leptons via a spin 0 mediator. In each of the four panels, the first two stacked plots show the product of mediator couplings to the DM and SM leptons and the annihilation cross section $\sigmav_{v\to0}$ to electrons against the mediator mass for masses and couplings that predict the correct relic density. There is a pronounced resonance at $2 m_\chi \approx M_\Phi$, at which annihilation proceeds through an $s$-channel resonance and a reduced coupling is required for the relic density. When the resonance is almost exactly on-shell in the low-velocity limit, it enhances $\sigmav_{v\to0} \gg \sigmav$ as the latter is reduced upon thermal averaging. Following the resonance, though, there is a region with reduced $\sigmav_{v\to0}$. In this region, in the low-velocity limit the resonance is off-shell but the thermally averaged cross section averages over the on-shell resonance, and thus $\sigmav_{v\to0} \ll \sigmav$. This region cannot explain the DAMPE result under our assumptions about the boost factor in Sec.~\ref{sec:dampe}.The points with an annihilation cross section large enough to explain the DAMPE excess with a boost factor of $17$ -- $35$ lie in the green region in the middle panels. Regions of mediator mass that are consistent with experimental data and could explain the DAMPE excess lie above and below the resonance (indicated by blue in the bottom panels).

The third stacked panel shows the strictest experimental constraint, which is DM direct detection (presently PandaX) for scalar interactions (left) and LEP searches for pseudoscalar interactions (right) between the mediator and SM leptons. Only the resonance region survives the DM direct detection constraint on the SI cross section for scalar interactions, as the $t$-channel interaction with quarks is not enhanced by the resonance. The resonance region, however, cannot produce a substantial $\sigmav_{v\to0}$ and is thus not of interest. For the pseudoscalar interaction, the DD constraints are weaker as there is no loop induced SI cross section and the SD cross section is momentum-suppressed, and LEP provides the strongest constraints. The survived region can be further probed in future DD experiments and $e^+e^-$ colliders. For example, the forthcoming LZ experiment\cite{Mount:2017qzi} will almost fully cover the scalar interaction cases, while ILC can detect mediators with a mass larger than $2m_{\chi}$ for all cases, with only a very small part of the resonance region that could escape.

In Fig.~\ref{fig:spin0_DMfermion}, we show the same story as in Fig.~\ref{fig:spin0_DM0} but for a spin 0 mediator interacting with Dirac and Majorana fermionic DM. The interaction between the fermionic DM and mediator is pseudoscalar as the annihilation cross section with a scalar interaction (not shown) is velocity suppressed and thus not of interest. The plots tell a similar story to a spin 0 mediator interacting with scalar DM, though in this case LEP is the most powerful constraint for scalar and pseudoscalar interactions between the mediator and SM leptons as SI and SD cross sections are momentum suppressed or absent.

We consider vector mediators coupled to fermionic DM in Fig.~\ref{fig:spin1_DMfermion}. The top row shows vector interactions between a Dirac fermionic DM particle and a vector mediator (this vector interactions is forbidden for Majorana fermions). Combined with a vector interaction between the mediator and SM leptons, we find unsuppressed SI cross sections and thus severe constraints from DM direct detection. The forthcoming DD experiments, such as XENONnT and LZ, will be sensitive to almost all of the small remaining viable region. For an axial-vector interaction, though, LEP constraints are most powerful as SI interactions are momentum suppressed. In the bottom row we show results for axial-vector interactions between fermionic DM and a vector mediator. In this case the annihilation cross section to fermions is helicity-suppressed, thus we see $\sigmav\approx10^{-30}\cmCubePerSec$, which is far too small, and annihilation to on-shell mediators ($Z^\prime Z^\prime$), which is unsuppressed, dominates if it is kinematically allowed. Although final-state radiation could lift the suppression, it would soften the posit on spectrum. The axial-vector interaction with DM and scalar with SM leptons (not shown) is velocity suppressed.

\begin{figure}[tbp]
\centering 
\includegraphics[width=.45\textwidth]{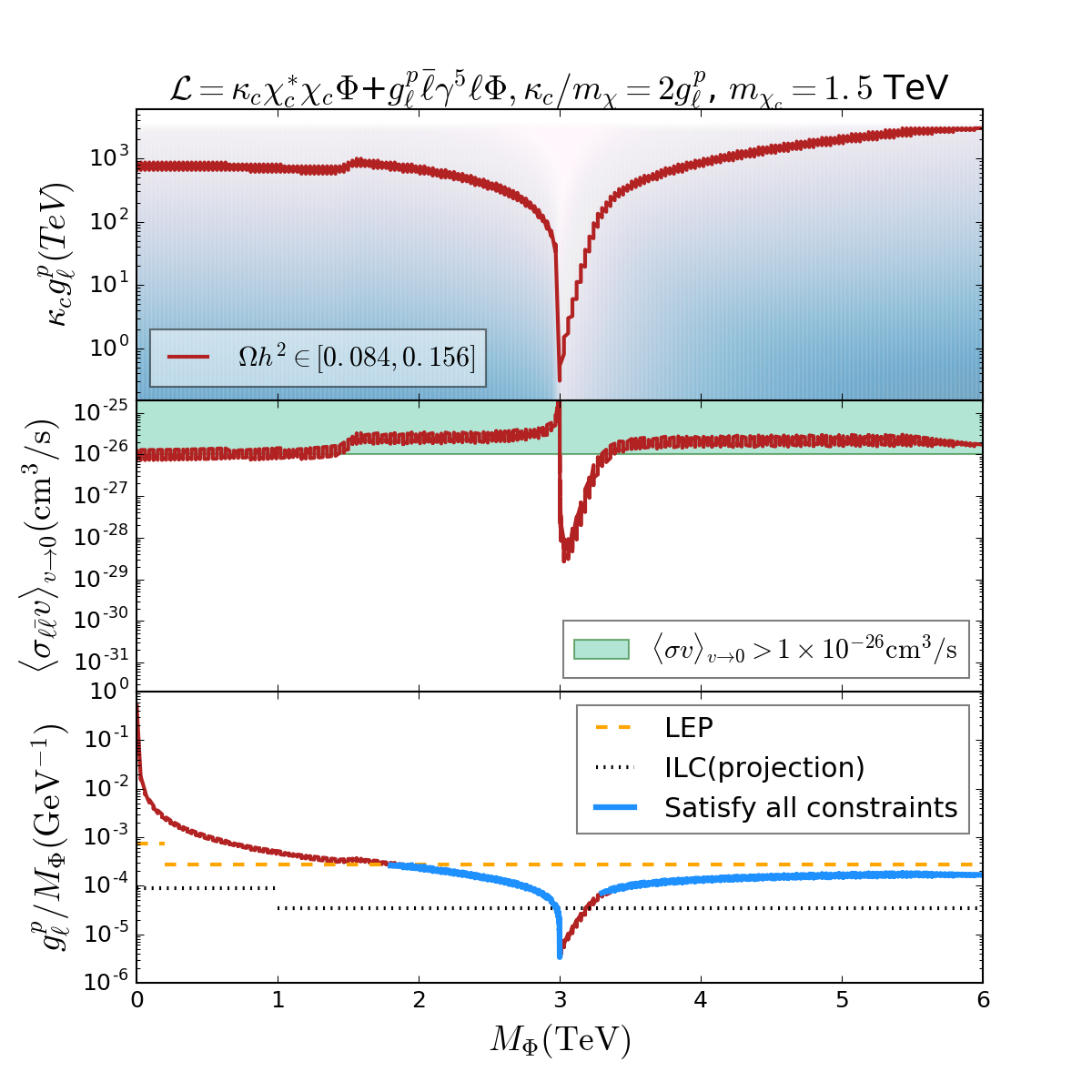}
\hfill
\includegraphics[width=.45\textwidth]{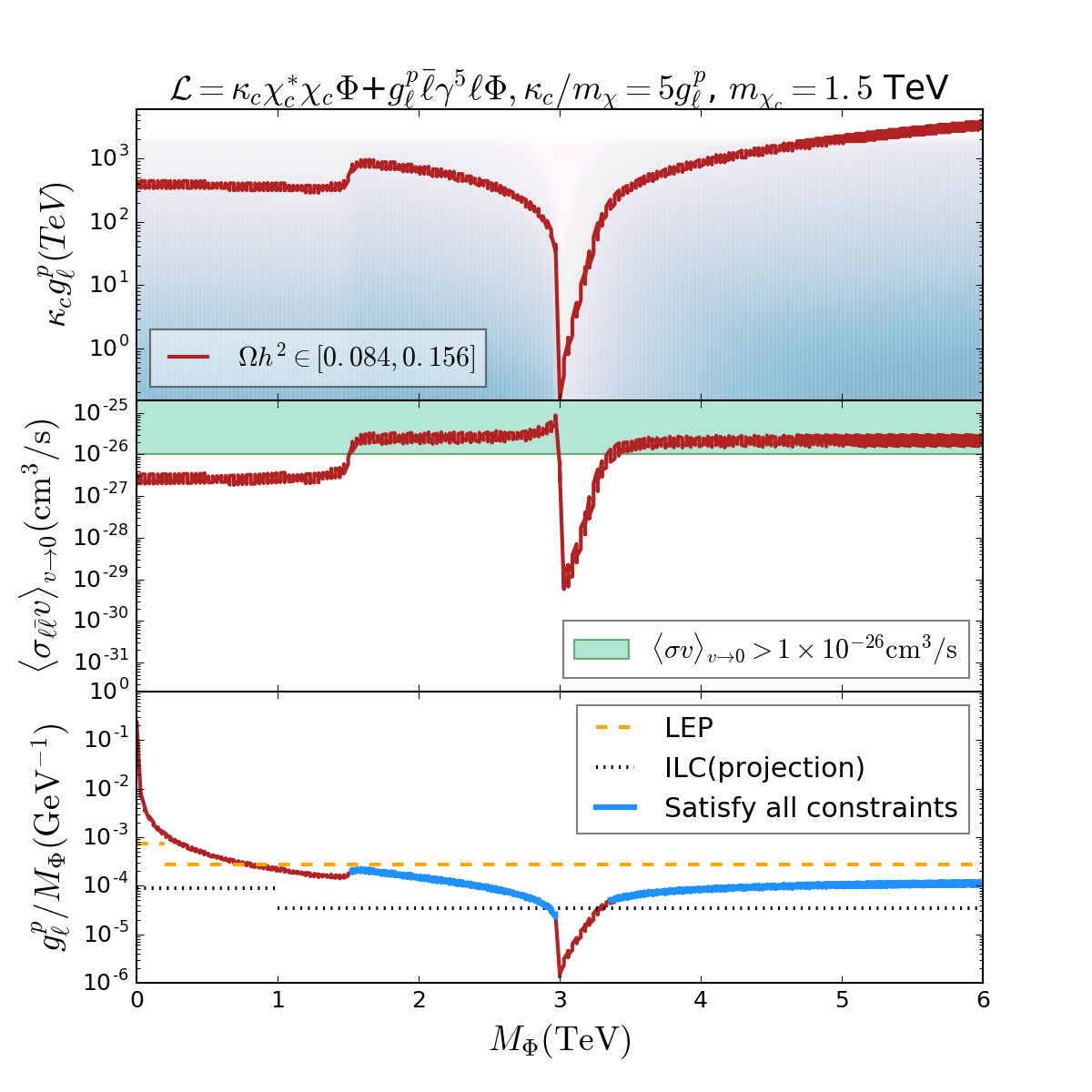}
\caption{\label{fig:lep_coupling} A scalar mediator scalar coupled to complex scalar DM and pseudoscalar coupled to leptons with $\kappa_c/m_{\chi}=2g_{\ell}^p$ (left) or $\kappa_c/m_{\chi}=5g_{\ell}^p$ (right). Each panel shows DM properties as a function of mediator mass. The blue color gradient in the first panel shows relic density, where darker blue means greater relic density.}
\end{figure}

In Fig.~\ref{fig:lep_coupling}, we relax our assumption that the mediator couplings to DM and SM leptons are equal for scalar mediator with a scalar interaction to scalar DM and pseudoscalar interaction to SM leptons. By decreasing the mediator coupling with SM leptons, we may evade LEP constraints. However, in doing so we reduce the rate at which DM annihilates to electrons. We see that once annihilation to pairs of on-shell mediators is kinematically impossible (at about $m_\Phi = 1.5\tev$), the required coupling slightly increases. This is most pronounced when we permit the coupling to leptons to be five times smaller (right) though visible when it is two times smaller (left). In the lowest stacked panels, we see that the LEP constraint is now considerably weaker than in Fig.~\ref{fig:spin0_DM0}.

\begin{figure}[tbp]
\centering 
\includegraphics[width=.45\textwidth]{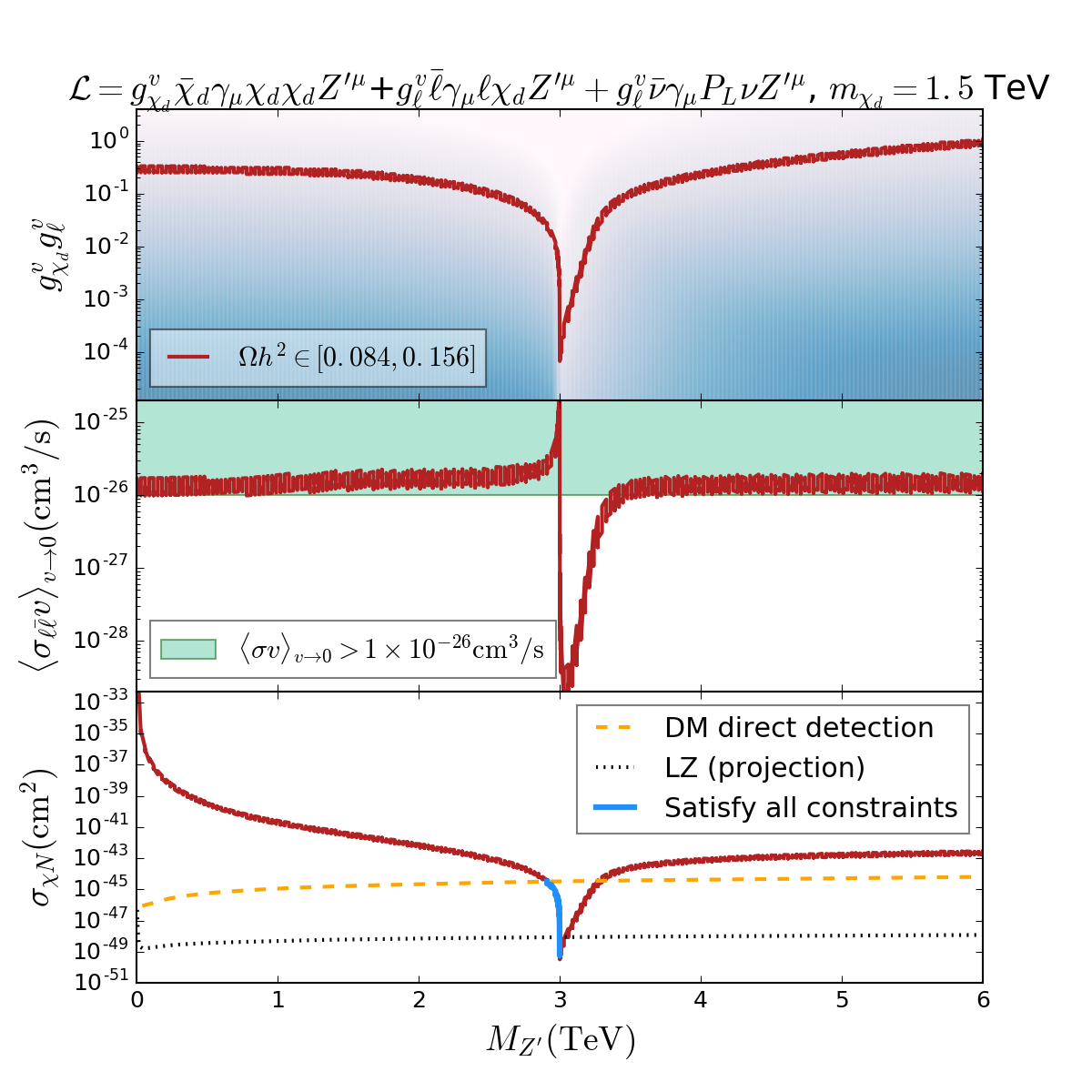}
\hfill
\includegraphics[width=.45\textwidth]{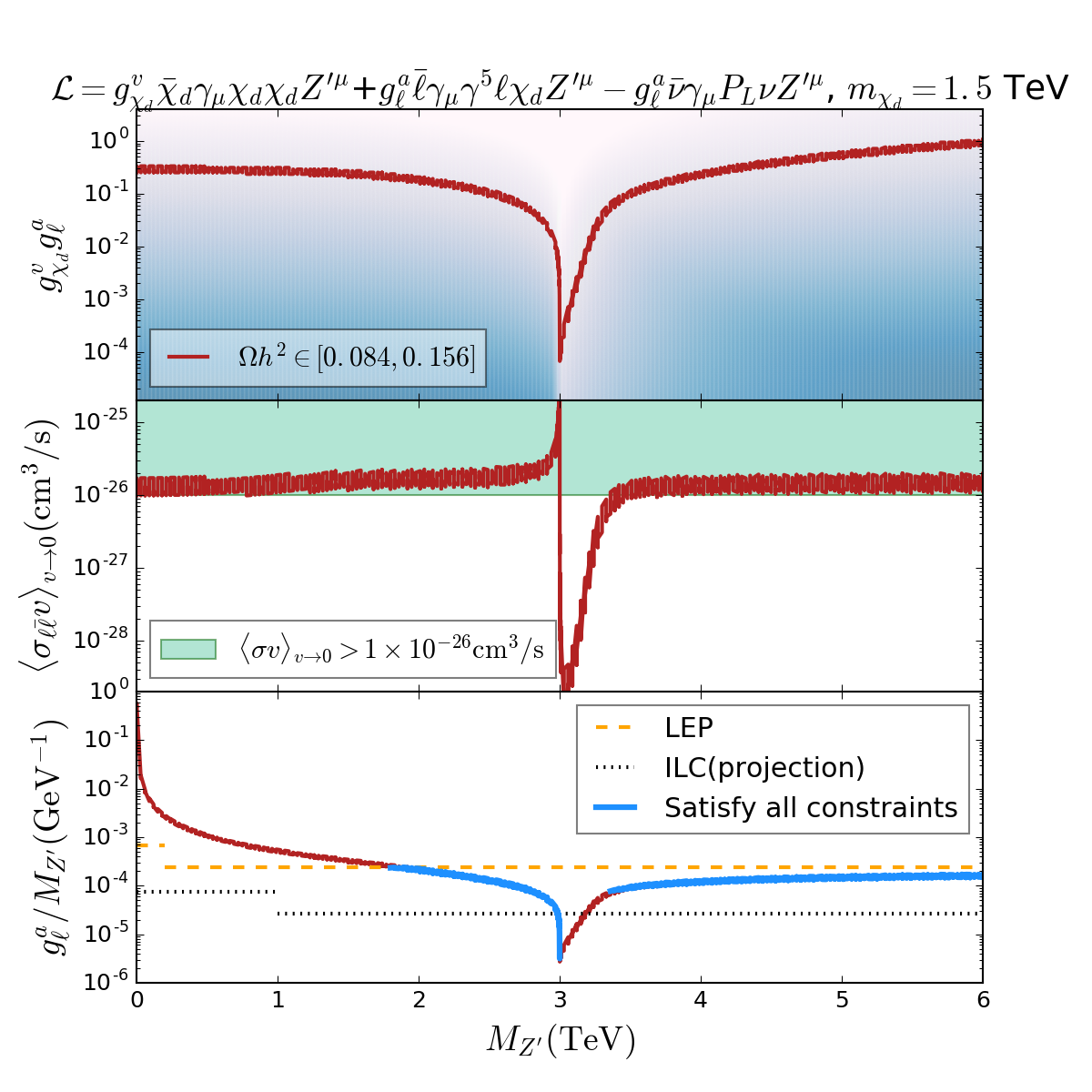}
\caption{\label{fig:neutrino} A vector mediator vector coupled to complex (left) and axial-vector (right) coupled to leptons and neutrinos simultaneously. Each panel shows DM properties as a function of mediator mass.  The blue color gradient in the first panel shows relic density, where darker blue means greater relic density.}
\end{figure}

The lepton current in Eq.~\ref{eq:la_lep} can be translated into a gauge invariant current associated with a left-handed weak isospin doublet and a right-handed isospin singlet, $g_{\ell}^{L/R}=g_{\ell}^v \mp g_{\ell}^a$. As a consequence, in any SU(2) invariant theory, the coupling between a vector mediator and neutrinos $g_{\nu}$ is in general non-zero, including in our scenarios with $g_{\ell}^v=0$ or $g_{\ell}^a=0$. Thus, in Fig.~\ref{fig:neutrino}, we show the cases of vector mediator with gauge invariant lepton current, i.e., $g_{\nu}=g_{\ell}^v$ if $g_{\ell}^a=0$ and $g_{\nu}=-g_{\ell}^a$ if $g_{\ell}^v=0$. Compared with Fig.~\ref{fig:spin1_DMfermion}, we see that the cross section of DM annihilating into leptons is slightly diluted by annihilation into neutrinos, but can still reach $10^{-26}\cmCubePerSec$. Although the constraints from trident production restrict the muon coupling if the mediator couples to muon neutrinos, the strictest limits in the $g_{\nu}=g_{\ell}^v$ and $g_{\nu}=g_{\ell}^a$ cases are PandaX and LEP, respectively, which are similar to the simplified cases.

\begin{table}[]\footnotesize
\centering
\begin{tabular}{cccccccc}
\toprule
$\mathcal{L}_\text{M--SM}$                         & $\mathcal{L}_\text{M--DM}$               & DAMPE signal              & $\Delta a_{\mu}$ & DM DD       & LEP  & Combined           \\
\midrule
$\ell\bar{\ell} \Phi$                              & $\chi_r\chi_r \Phi$                      & $\notin [3001,3330]$ & $>361$   & $\in[2971,3240]$  & $>2010$ &$\in[2971,3001]$ \\
$\ell\bar{\ell} \Phi$                              & $\chi_c^*\chi_c \Phi$                    & $\notin [3002,3270]$ & $>420$   & $\in[2970,3180]$  & $>2160$ &$\in[2970,3002]$\\
$\ell\bar{\ell} \Phi$                              & $\chi_d\gamma^5\chi_d \Phi$              & $\notin [3001,3270]$ & $>361$   & $>276$            & $>2010$    &$\notin [3001,3270]\&>2010$\\
$\ell\bar{\ell} \Phi$                              & $\chi_m\gamma^5\chi_m \Phi$              & $\notin [3001,3331]$ & $>301$   & $>241$            & $>1770$ &$\notin [3001,3331]\&>1770$\\
$\ell\gamma^5\bar{\ell} \Phi$                      & $\chi_r\chi_r \Phi$                      & $\notin [3001,3300]$ & $>391$   & --             & $>1980$ &$\notin [3001,3300]\&>1980$\\
$\ell\gamma^5\bar{\ell} \Phi$                      & $\chi_c^*\chi_c \Phi$                    & $\notin [3002,3270]$ & $>450$   & --             & $>2160$ &$\notin [3002,3270]\&>2160$\\
$\ell\gamma^5\bar{\ell} \Phi$                      & $\chi_d\gamma^5\chi_d \Phi$              & $\notin [3001,3270]$ & $>420$   & --             & $>2010$ &$\notin [3001,3270]\&>2010$\\
$\ell\gamma^5\bar{\ell} \Phi$                      & $\chi_m\gamma^5\chi_m \Phi$              & $\notin [3001,3331]$ & $>330$   & --             & $>1770$ &$\notin [3001,3331]\&>1770$\\
\midrule
$\bar{\ell}\gamma_{\mu}\ell Z^{\prime\mu}$         & $\chi_d\gamma_{\mu}\chi_d Z^{\prime\mu}$ & $\notin [3000,3330]$ & $>61$    & $\in[2940,3270]$  & $>2130$ &$\in [2940,3000]$\\
$\bar{\ell}\gamma_{\mu}\gamma^5\ell Z^{\prime\mu}$ & $\chi_d\gamma_{\mu}\chi_d Z^{\prime\mu}$ & $\notin [3000,3330]$ & $>270$   & --             & $>1890$&$\notin [3000,3330]\&>1890$\\
\bottomrule
\end{tabular}
\caption{Regions in mediator mass permitted by experimental constraints for combinations of mediator couplings to the SM leptons and DM. 
The constraint DM DD stands the spin-independent DM direct detection constraints currently provided by PandaX. The final column shows mediator masses that may explain DAMPE and simultaneously satisfy all experimental constraints. All masses are in units of \gev.}
\label{tab:mass}
\end{table}

We do not show vector mediators coupling to scalar DM as $\sigmav_{v\to0}$ is velocity suppressed. We summarize the regions of mediator mass that are excluded in Table~\ref{tab:mass}.  There one can see that the most important constraints come from LEP and DM direct detection (presently PandaX), along with a small constraint from DAMPE in the resonance region. The region where one can explain the deviation between the SM calculation and experiment for the anomalous magnetic moment of the muon is entirely excluded by the LEP limit.  At the same time the limit from the more conservative assumption about theory errors discussed in section \ref{sec:amu} is much weaker than the LEP limit.  For simplicity we only show the latter in Table~\ref{tab:mass}.  An overview of our findings are presented Table~\ref{tab:sum}.  This shows the combinations of operators with velocity or helicity suppressed $\sigmav$, which are therefore not of interest, other combinations with unsuppressed scattering cross sections that are in tension with DD experiments, and finally that for the remaining models LEP has the largest impact, but that these models are still allowed.

\section{Conclusions}
\label{sec:conclusions}
We performed a model-independent analysis of particle dark matter explanations of the peak in the DAMPE electron spectrum and whether they can simultaneously satisfy constraints from other DM searches. We assumed that the signal originated from DM annihilation in a nearby subhalo with an enhanced density of DM. To account for the inevitable energy loss, we assumed a DM mass of about $1.5\tev$, which is slightly greater than the location of the observed peak. Rather than working in a specific UV-complete model, we investigated all renormalizable interactions between SM leptons, DM of spin 0 and 1/2, and mediators of spin 0 and 1. We did not consider the more exotic cases of spin-1 DM, a spin-1/2 $t$-channel mediator or a spin-2 mediator in this work. 

Our results are summarized in Tables~\ref{tab:sum} and \ref{tab:mass}. We found that 10 of 20 possible combinations of operators are helicity or velocity suppressed and cannot explain the DAMPE signal assuming a nearby subhalo with a density enhancement of $17$ -- $35$, as calculated in\cite{Yuan:2017ysv}. Of the remaining combinations, DM direct detection strongly constrains the unsuppressed scattering cross sections in three models and LEP strongly constrains the mass of the mediator in the other 7. The remaining candidates are (1) a spin 0 mediator coupled to scalar DM, (2) a spin 0 mediator pseudoscalar coupled to fermionic DM, and (3) a spin 1 mediator vector coupled to Dirac DM. LEP constraints on four-fermion operators force the mediator mass to be heavy, $\gtrsim 2\tev$, in all of these scenarios.

\acknowledgments

This research, in part, was supported by the ARC Centre of Excellence for Particle Physics at the Tera-scale, grant CE110001004. The work of PA was also supported by Australian Research Council grant, FT160100274.

\bibliographystyle{JHEP}
\bibliography{DAMPE}

\end{document}